\documentclass[10pt]{iopart}
\usepackage{iopams,bm,epsfig}

\newcommand{\beas}{\begin{eqnarray*}}
\newcommand{\eeas}{\end{eqnarray*}}
\newcommand{\bea}{\begin{eqnarray}}
\newcommand{\eea}{\end{eqnarray}}

\newcommand{\beq}{\begin{equation}}
\newcommand{\eeq}{\end{equation}}
\newcommand{\half}{{\textstyle{1\over2}}}

\begin{document}
\title{Spin dynamics of low-dimensional excitons due to acoustic phonons}
\author{A. Thilagam and M. A. Lohe}
\address{Department of Physics, The University of Adelaide, Australia 5005}
\ead{Max.Lohe@adelaide.edu.au}

\begin{abstract}
We investigate the spin dynamics of excitons
interacting with acoustic phonons in quantum wells,  
quantum wires and quantum disks by employing a multiband
model based on the $4\times4$ Luttinger Hamiltonian.
We also use the Bir-Pikus Hamiltonian to model the coupling of excitons
to both longitudinal acoustic phonons and transverse acoustic phonons,
thereby providing us with a realistic framework  in which to
determine details of the spin dynamics of excitons.
We use a fractional dimensional
formulation to model the excitonic wavefunctions and we demonstrate explicitly the 
decrease of spin relaxation time with dimensionality.
Our numerical results are consistent with 
experimental results of spin relaxation times for various
configurations of the GaAs/Al$_{0.3}$Ga$_{0.7}$As material system. 
We find that longitudinal and transverse acoustic phonons are 
equally significant in processes of exciton spin relaxations
involving acoustic phonons. 

\end{abstract}



\section{Introduction}

Excitons in  low dimensional systems,  
formed due to the binding of 
electrons and holes, have  well known features such as enhanced
binding energies and oscillator strengths \cite{vin}.
In recent years there have been increased studies of excitonic 
processes involving
phonons within the restricted dimensional 
exciton energy band \cite{haug,shah}, and exciton spin relaxation \cite{mai},  
formation and phonon-assisted thermalization \cite{oh,sok,iva} 
have attracted great interest. This is largely due
 to the significant advancement in spectroscopic 
techniques involving continuous-wave and time-resolved photoluminescence
excitation measurements \cite{mei} which provide results of high 
temporal
and spatial resolution. In particular, the 
use of polarization dependent photoluminescence \cite{da1,da2} involving
circularly polarized light has extended the possible application
of excitonic systems to spintronics \cite{wolf,buyanova} 
in which the spin degree of freedom is exploited in electronic devices.

Several processes involving the Elliot-Yafet 
mechanism, the Bir-Aroniv-Pikus mechanism 
and the Dyakonov-Perel mechanism 
are known to cause spin relaxation of carriers
in solids \cite{tit}. We are interested in the first
of two other possible mechanisms which can cause 
exciton scattering involving spin-flip processes,
firstly, holes interacting with acoustic phonons \cite{mai2,Une,jap1}
and, secondly, long-range electron-hole exchange interactions
which couple the spins of electrons and holes \cite{s}.
In narrow wells both mechanisms can become comparable,
the dominant process depending on the electron density, 
temperature and quantum well width. 

Spin dynamics of excitons are identified through measurements
of circular polarization \cite{mei}-\cite{da2}
in which, initially, spin polarized electrons
and holes are formed followed by both spin and energy relaxation of carriers and,
finally, there is recombination of carriers 
resulting in the measurement of a circularly polarized signal. It is not
clear, however,  whether the electron and hole undergo spin and
energy relaxation as unlinked carriers and then become coupled
(forming an exciton) after which they recombine, or whether
the carriers become coupled first, then undergo spin and
energy relaxation before  the recombination process. 
Whether there is concurrence of both spin 
and energy relaxation in electron-hole systems also 
needs careful study. 
There have been suggestions \cite{jap1} that spin relaxation 
occurs after energy relaxation into
the excitonic ground state, while others \cite{snoke}
have suggested that  momentum relaxation
occurs simultaneously with spin relaxation
in excitonic systems. This interplay 
of energy and spin relaxation processes  
needs further study in order to gain a thorough understanding
of phonon related dynamics in low dimensional
semiconductor systems. Such details are also vital
for the accurate interpretation of experimental 
results. 

In a simple exciton model the fourfold degeneracy at the top of the valence band, 
which introduces some nontrivial features in the excitonic spectrum, is neglected. 
The complicated nature of the multiband Wannier exciton model was noted
by Dresselhaus \cite{dress} who demonstrated the absence of a well-defined exciton 
centre-of-mass transformation.  
Subsequently Baldereschi and Lipari \cite{alte1,alte2}
studied  exciton dispersion relations for
a variety of semiconductor systems using 
a generalized centre-of-mass transformation.
In the Baldereschi-Lipari approach, the  ``$d$-like" term is treated as a 
perturbation on the ``$s$-like" term of the 
exciton Hamiltonian which has a hydrogen-like spectrum.
Using this approach, Kane \cite{kane1} obtained  useful expressions for the energy
shift of the bulk $1s$ exciton from the conduction
band minimum.  In recent years, the multiband model of the 
exciton interacting with phonons in bulk semiconductor systems,
investigated in \cite{tre,beni}, 
has been extended to the quasi-two dimensional model of the 
exciton \cite{Une,sia,rudin}. 
These papers show that the description of low-dimensional excitons 
within a multiband model is a complex problem that
requires extensive computational efforts.  
Here we use a fractional dimensional formalism to extend the study of spin dynamics
of excitons to quantum wires and quantum disks and to
emphasize the significance of the 
effect of dimensionality in the spin dynamics of excitons.

The paper is organized as follows. In Section~\ref{sec:theo} 
we present the theoretical formulation of the quasi-two dimensional
exciton model that is used to obtain suitable dispersion relations 
as well as an explicit form of the exciton wavevector.
In Section~\ref{sec:frac} we give details 
of the fractional dimensional approach and present
results specific to the quantum wire system.
In Section~\ref{sec:phonon} we  use the Bir-Pikus 
Hamiltonian to model the coupling of excitons
to acoustic phonons and obtain expressions
of  matrix elements corresponding to longitudinal acoustic (LA) and 
transverse acoustic (TA)  phonon modes.
In Section~\ref{sec:matrix}
we extend the theory developed in Sections \ref{sec:theo}
and \ref{sec:phonon} to obtain expressions
 of spin transitions corresponding to
the decay of heavy- and light-hole excitons
and  intraband as well as interband scattering 
in constrained systems. We report and discuss the results of 
salient features of excitonic processes 
mediated by LA and TA phonons in Section~\ref{sec:discuss}, and
summarize our conclusions in Section~\ref{sec:concl}.


\section{Theoretical Formulation}
\label{sec:theo}

We investigate properties of Wannier excitons in quantum wells
fabricated using direct-gap cubic semiconductors, using as an 
example GaAs which is representative of a large
class of III-V and II-VI zinc blende semiconductors.
The conduction band edge in such systems involves an 
$s-$type spin degenerate 
carrier state while the valence band edge involves a $p-$symmetry 
state with a sixfold degeneracy \cite{haug,kane}.
When spin-orbit interactions are included the sixfold
degenerate level $\Gamma_5$ splits into a fourfold degenerate state
$\Gamma_8$ (with $J={3\over2}$) and a twofold degenerate state $\Gamma_7$
($J={1\over2}$). We neglect the twofold degenerate state, also known
as the splitoff band, in this paper. 
When analysing these spin states, we use  a basis  that is 
oriented relative to the quantization axis, where the $Z$-axis is
chosen in the growth direction of the quantum well structure. 
The choice of a preferential axis of 
quantization for the hole spin subsequently leads to notable anisotropic
properties of the exciton as we show later in this work. 

The vectors in the angular momentum basis generated by
${\bf J}=\left(J_{x},J_{y},J_{z}\right)$
projected on the quantization axis are labelled by the magnetic quantum
numbers $m = \pm \frac{3}{2}$ for the heavy
hole (HH) states and  $m = \pm\frac{1}{2}$ for the light- hole 
(LH) states, for the degenerate level $\Gamma_8$ (see Chapter 3 in 
\cite{shah}).
The broken translational symmetry in the growth direction of quantum wells
lifts the degeneracy of the heavy and light-hole valence bands
at the $\Gamma$-point, giving rise to a distinct splitting
at the two-dimensional in-plane hole wavevector $\vec{k}_{{\rm h}} = 0$.
This is the basis of the formation of heavy- and light-hole excitons 
in quantum wells. At $\vec{k}_{{\rm h}} \ne 0$, the LH and HH states mix and hybridize
giving rise to strong nonparabolicity \cite{faso,bro} in the valence
band spectrum. Due to the distortion of the valence band structure,
 we classify the heavy- and light-hole exciton bands 
by their basis functions at $\vec{k}_{{\rm h}} = 0$.  
Unlike the split-off band which arises from the spin-orbit coupling,
the splitting between the heavy-hole and light-hole states is a 
consequence of strain due to confinement in the growth direction.
Throughout this paper
$\vec{k}_{{\rm e}}$ ($\vec{k}_{{\rm h}}$) marked with an arrow denotes the 
in-plane electron (hole) wavevector, while 
$\bf{k_{{\rm e}}}$ ($\bf{k_{{\rm h}}}$) denotes the three-dimensional
electron (hole) wavevector. 

In order to define operators associated with the spins of charge carriers
in quantum wells we use the standard rotation matrix
$D(\alpha,\beta,\gamma)$, which is a function
of the Euler angles (where 
$0 \leqslant \alpha \leqslant 2 \pi, 
0 \leqslant \beta \leqslant \pi, 0 \leqslant \gamma \leqslant 2 \pi$),
and is given by \cite{edmond,brink}:
\[
D(\alpha,\beta,\gamma)
=
\exp(-\rmi \alpha J_z)\exp(-\rmi \beta J_y) \exp(-\rmi \gamma J_z).
\]
The angles $\alpha = \phi$
and $\beta = \theta$ 
are in fact the polar angles of the  spin state $\hat{J}(\theta,\phi)$
with respect to the frame in which $Z$ is 
the quantization axis. A  rotated spin state 
can be represented as a combination of the
$2J+1$ component states using standard rotation matrices:
\beq
\label{eq:rotcomb}
D(\alpha,\beta,\gamma) | J m \rangle =
\sum_{m'} \langle J m^\prime| 
D(\alpha,\beta,\gamma)|J m \rangle \;
|Jm^\prime \rangle .
 \eeq
The matrix elements $D_{m'm}^J (\alpha,\beta,\gamma) = 
\langle J m'| D(\alpha,\beta,\gamma)|
J m \rangle$, where $-J\leqslant m,m' \leqslant J$,
may be written out explicitly for any $J$, however we require only
the explicit form and transformation properties for angular momentum
values $J=\frac{1}{2}$ and $J=\frac{3}{2}$ which
are applicable to the electron and
hole.  Details are given by Brink and Satchler \cite{brink}. 
As the basis vectors of the representation in equation (\ref{eq:rotcomb})
are chosen as the eigenfunctions of $J_z$,  we use the simplification
$|{\it D}_{m'm}^J (\alpha,\beta,\gamma)| = 
|d_{m'm}^J(\beta)|$ where $d^J$ is the 
reduced rotation matrix. A variety of relations 
among the matrix elements $d_{m'm}^J (\beta)$ 
is given in Chapter 3 of the monograph by Biedenharn and Louck \cite{BL}.

\subsection{Electron and hole operators with arbitrary spin orientation.}
\label{sec:theo1}

By using properties of the reduced rotation matrices we are able to
write the  quasi-two dimensional electron and heavy- (or light-hole) creation operators 
as linear combinations of operators corresponding to the 
spin projections on the quantization axis. 
The creation operators of the electron 
$\hat a_{\vec{k}_{{\rm e}}}^{\dagger}({\varsigma})$, 
heavy hole $\hat c_{\vec{k}_{{\rm h}}}^{\dagger}({\eta})$ 
and light-hole $\hat c_{\vec{k}_{{\rm h}}}^{\dagger}(\eta)$ spinor systems
thus appear as follows:
\bea
\label{eq:creop}
\hat a_{\vec{k}_{{\rm e}}}^{\dagger}({\varsigma}) &=& \sum_{\tau_z} 
\; d_{\varsigma  \frac{1}{2} \; \tau_z}^{1/2}(\theta_{{\rm e}})\;
\hat a_{\vec{k}_{{\rm e}}}^{\dagger}({\tau_z}), \nonumber \\
\hat c_{\vec{k}_{{\rm h}},{\rm H}}^{\dagger}({\eta}) &=& \sum_{\sigma_{z}}
\;  d_{ \eta \frac{3}{2} \; \sigma_{z}}^{3/2}(\theta_{{\rm h}})\;
\hat c_{\vec{k}_{{\rm h}}}^{\dagger}({\sigma_{z}}), \nonumber \\
\hat c_{\vec{k}_{{\rm h}},{\rm L}}^{\dagger}(\eta) &=& \sum_{\sigma_{z}} 
\; d_{\eta \frac{1}{2} \; \sigma_{z}}^{3/2}(\theta_{{\rm h}})\;
\hat c_{\vec{k}_{{\rm h}}}^{\dagger}({\sigma_{z}})
\eea
where $\tau_z = \pm \frac{1}{2}$ and $\sigma_z = \pm \frac{3}{2} 
(\pm \frac{1}{2})$ are the respective electron and 
heavy (light-) hole spin projection eigenvalues of $J_z$.
The indices $\varsigma = \pm $ and  $\eta = \pm $ denote the 
two possible  degenerate states of the electron (at $\vec{k}_{{\rm e}} = 0$)
and heavy and light-holes (at $\vec{k}_{{\rm h}} = 0$) respectively, and
$\theta_{{\rm e}}$ ($\theta_{{\rm h}}$) denotes the  azimuthal 
angle of spin states of the electron (hole) operators
with the in-plane wave vector ${\vec{k}_{{\rm e}}}$ (${\vec{k}_{{\rm h}}}$).
For simplicity, we use the same notation  for 
the heavy- and light-hole wavevectors and the corresponding azimuthal angles. 

The dependence of the operators  $\hat a_{\vec{k}_{{\rm e}}}^{\dagger}({\varsigma})$,
$\hat c_{\vec{k}_{{\rm h}},{\rm H}}^{\dagger}({\eta})$ and  
$\hat c_{\vec{k}_{{\rm h}},{\rm L}}^{\dagger}({\eta})$ ) on their respective
azimuthal angles yields the expected relations at selected 
angles. For instance,  $\theta_{{\rm e}} =  0$ yields the relations: 
$\hat a_{\vec{k}_{{\rm e}}}^{\dagger}(+) = 
\hat a_{\vec{k}_{{\rm e}}}^{\dagger}(\frac{1}{2})$ and
$\hat a_{\vec{k}_{{\rm e}}}^{\dagger}(-) =
\hat a_{\vec{k}_{{\rm e}}}^{\dagger}(-\frac{1}{2})$.
Likewise $\theta_{{\rm e}} =  \pi$ yields the relations: 
 $\hat a_{\vec{k}_{{\rm e}}}^{\dagger}(+) = 
- \hat a_{\vec{k}_{{\rm e}}}^{\dagger}(-\frac{1}{2})$ and
$\hat a_{\vec{k}_{{\rm e}}}^{\dagger}(-) =
-\hat a_{\vec{k}_{{\rm e}}}^{\dagger}(\frac{1}{2})$. 
Similar relations can be obtained
for the two types of hole spinor systems.


\subsection{Exciton Wavefunction and Spin}
\label{sec:theo2}

The field operator $\hat\Psi^\dagger_\sigma$
of an exciton, with a centre of mass
that moves freely, is  composed of 
electron and hole operators with spins aligned along
the quantization axis, and has the expression:
\begin{eqnarray}
\label{eq:field}
\hat\Psi^\dagger_\sigma({\bf{r_{{\rm e}}},{r_{{\rm h}}}})
=
 \sum_{\vec{k}_{{\rm e}}, \vec{k}_{{\rm h}}} 
&\left[
\xi_{\vec{k}_{{\rm e}}}(z_{{\rm e}}) \; 
\e^{\rmi\vec{k}_{{\rm e}}\cdot \vec{r_{{\rm e}}}} 
\sum_{\tau_z} 
u_{\bf{k_{{\rm e}}}}^{\tau_z}({\bf{r_{{\rm e}}}})\; 
a^\dagger_{\vec{k}_{{\rm e}}}(\tau_z)
a_{\vec{k}_{{\rm e}}}(\tau_z) 
\right.
\\
\nonumber
&+
\left. 
\sum_{\sigma_z} \; \xi_{\vec{k}_{{\rm h}}}^{\sigma_z}(z_{{\rm h}})\;
 \e^{- \rmi\vec{k}_{{\rm h}}\cdot \vec{r}_{{\rm h}}}\;
 u_{-\bf{k_{{\rm h}}}}^{-\sigma_z}({\bf{r_{{\rm h}}}})
c_{\vec{k}_{{\rm h}}}(\sigma_z)
c_{\vec{k}_{{\rm h}}}^\dagger(\sigma_z)
\right],
\end{eqnarray}
where ${\bf{r_{{\rm e}}}}=\left(x_{{\rm e}},y_{{\rm e}},z_{{\rm e}}\right)$
and ${\bf{r_{{\rm h}}}} =\left( x_{{\rm h}},y_{{\rm h}},z_{{\rm h}}\right)$
denote the space coordinates of the electron and hole, respectively, and
$\vec{r_{{\rm e}}}$ 
(or $\vec{r_{{\rm h}}}$) marked with an arrow denotes the in-plane 
coordinates of the electron (or hole). 
$u_{{\bf{k_{{\rm e}}}}}^{\tau_z}({\bf{r_{{\rm e}}}})$ 
($u_{{-\bf{k_{{\rm h}}}}}^{-\sigma_z}({\bf{r_{{\rm h}}}})$) is the  
periodic component of the
bulk Bloch function of the electron (hole) which 
depends on the projection $\tau_z$ ($\sigma_z$) of the 
spin momentum onto the positive $Z$-axis.  
As usual we assume
that the Bloch functions are identical
in both well and barrier materials. The conduction and valence bands in 
semiconductors with zinc blende structures 
are made up from $s$- and $p$- type atomic functions respectively.
Denoting now by $|U\rangle$ the orbital component of the 
$s$-type Bloch functions, we write
the electron Bloch wavefunctions as:
 \bea
\label{eq:bloche}
u_{\bf{k_{{\rm e}}}}^{1/2}({\bf{r_{{\rm e}}}}) &=& |\half,\half\rangle 
= |U \! \uparrow  \rangle , \nonumber \\
u_{\bf{k_{{\rm e}}}}^{-1/2}({\bf{r_{{\rm e}}}}) &=& |\half,-\half\rangle 
= |U \! \downarrow  \rangle,
\eea
where as usual $\uparrow, \downarrow$ denote spin-up and spin-down states.
Likewise, by denoting the Bloch functions with the 
symmetry of $p_x$, $p_y$, and $p_z$
orbitals as $|X\rangle $, $|Y\rangle$, and $|Z\rangle$, 
respectively, we may write the hole Bloch wavefunctions 
as linear combinations 
of $|(X+\rmi Y) \! \downarrow \rangle$, $|(X+\rmi Y) \! \uparrow \rangle$,
$|(X-\rmi Y) \! \downarrow \rangle$, $|(X-\rmi Y) \! \uparrow \rangle$,
$| Z \! \downarrow \rangle$ and  $|Z \! \uparrow \rangle$. 
 
It is well known that the  conservation of the in-plane wavevectors 
$\vec{k}_{{\rm h}}$ and $\vec{k}_{{\rm e}}$ in quantum wells allows the lateral 
motion of the charge carriers to be decoupled from 
their motion along the $Z$-axis. This gives rise to  
subbands of quantized envelope functions,
$\xi_{\vec{k}_{{\rm e}}}(z_{{\rm e}})$
and $\xi_{\vec{k}_{{\rm h}}}^{\sigma_z}(z_{{\rm h}})$,
in the conduction and valence bands respectively.
The electron subband function which has even or odd parity under $z_{{\rm e}}$
inversion is determined by using  the 
eigenfunction of the BenDaniel-Duke model \cite{duke}
in which the potential is 
approximated using a finite square well.
The boundary conditions at the interfaces 
are determined by using the wavefunction and 
current continuity  relations and 
\beq
\label{eq:holesubband1}
\xi_{\vec{k}_{{\rm e}}}(z_{{\rm e}}) =  
\sum_{\tau_z=\pm{1\over2}} \; \e^{\rmi\tau_z 
\phi_{{\rm e}}}\,\xi_{|\vec{k}_{{\rm e}}|}^{\tau_z}(z_{{\rm e}}),
\eeq
where $\vec{k}_{{\rm e}} = (k_{{\rm e}},\phi_{{\rm e}})$.

Calculation of the hole envelope function
$\xi_{\vec{k}_{{\rm h}}}^{\sigma_z}(z_{{\rm h}})$ is not 
as straight forward, as its dependence on 
$\vec{k}_{{\rm h}}$ and spin components $\sigma_z$ gives rise to a 
complicated warping of the various
valence subbands \cite{bro}. The axial approximation, in which
a cylindrical symmetry around the crystal 
growth direction is assumed (following \cite{Une}),
simplifies the hole subband states which may be expressed as a superposition of
four states of the valence band top with spin 
projections $\sigma_z = \textstyle{3\over2}, \half,-\half,-\textstyle{3\over2}$:
\beq
\label{eq:holesubband}
\xi_{\vec{k}_{{\rm h}}}(z_{{\rm h}}) =  
\sum_{\sigma_z} \; \e^{\rmi\sigma_z 
\phi_{{\rm h}}}\,\xi_{|\vec{k}_{{\rm h}}|}^{\sigma_z}(z_{{\rm h}})
\eeq
where $\vec{k}_{{\rm h}} = (k_{{\rm h}},\phi_{{\rm h}})$. 
Further details of the evaluation of the hole wavefunctions,  
characterized by the Luttinger parameters 
$\gamma_1,\gamma_2$ and $\gamma_3$, are given in 
the Appendix.

An exciton carrying
a centre-of-mass momentum ${\vec{K}}$ is also
characterized by its spin $S$ and its projection $S_{z}$
on the quantization axis. A given exciton spin state depends
on the  spin-spin correlations which exist 
between the electron and hole spinor systems. We consider
 only the $1s$ exciton so that  
the exciton spin $S$ is determined by the  group theoretical rules of 
angular momentum vector addition. 
Thus the four-dimensional state space of the exciton system 
spanned by the states $|S\rangle$ is characterized by a coupled system 
comprising the two-dimensional electron ($|\tau_z \rangle$)
and hole ($|\sigma_z \rangle$) spin states.
In this paper we choose the exciton basis states
with respect to the same axis 
of quantization as that of the hole and electron.
 
We define a {\it generalized} exciton as one that involves the coupling of 
an electron with wavevector ${\vec{k}_{{\rm e}}}$ and spin orientation angle
$\theta_{{\rm e}}$ to a heavy- or  light-hole with wavevector $ {\vec{k}_{{\rm h}}}$ 
and spin orientation angle $\theta_{{\rm h}}$. 
At low densities the exciton creation and annihilation
operators satisfy boson commutation relations, but
at higher densities we write the exciton 
eigenvector $|{\vec{K}}; S \rangle$ explicitly as a linear combination of 
all products of possible pairs of electron and hole eigenvectors: 
\begin{eqnarray}
\fl
\label{eq:exstate}
|{\vec{K}};S  \rangle 
\\
\fl
\nonumber
= 
\sum_{\stackrel{\vec{k}_{{\rm e}},\vec{k}_{{\rm h}}}{\varsigma,\eta}}
\left[
F_{{\rm H}}(\vec{k}_{{\rm e}},\vec{k}_{{\rm h}};\varsigma,\eta)\;
\left|\vec{k}_{{\rm e}},\vec{k}_{{\rm h}};\varsigma,\eta  \right\rangle_{{\rm H}}
+ F_{{\rm L}}(\vec{k}_{{\rm e}},\vec{k}_{{\rm h}};\varsigma,\eta)\; 
\left|\vec{k}_{{\rm e}},\vec{k}_{{\rm h}};\varsigma,\eta \right\rangle_{{\rm L}} 
\right]
\; \delta_{\vec{K}, \vec{k}_{{\rm e}} - \vec{k}_{{\rm h}}}
\end{eqnarray}
where $F_{{\rm H}}({\vec{k}_{{\rm e}}},{\vec{k}_{{\rm h}}};\varsigma,\eta)$ 
($F_{{\rm L}}({\vec{k}_{{\rm e}}},{\vec{k}_{{\rm h}}};\varsigma,\eta)$) represents the
coefficient function of coupling between the
electron and  heavy-hole (light-hole). We have omitted the 
azimuthal angles 
$\theta_{{\rm e}}$ and $\theta_{{\rm h}}$ of the spin states 
in equation (\ref{eq:exstate}), for simplicity in notation. 
The eigenvectors $|{\vec{k}_{{\rm e}}},{\vec{k}_{{\rm h}}};\varsigma,\eta \rangle_{{\rm H}}$ 
and $|{\vec{k}_{{\rm e}}},{\vec{k}_{{\rm h}}};\varsigma,\eta \rangle_{{\rm L}}$ 
are constructed from linear combinations of the 
electron and hole creation operators of the respective forms: 
$\hat a_{\vec{k}_{{\rm e}}}^\dagger({\varsigma}) \; 
\hat c_{\vec{k}_{{\rm h}},{\rm H}}^{\dagger}({\eta}) |0 \rangle$ and 
$\hat a_{\vec{k}_{{\rm e}}}^\dagger({\varsigma}) \; 
\hat c_{\vec{k}_{{\rm h}},{\rm L}}^{\dagger}({\eta})|0 \rangle$, 
where $|0 \rangle$ is the vacuum state.
We  have considered only the lowest heavy-hole and light-hole 
exciton states and therefore neglected energy band indices in 
equation (\ref{eq:exstate}). The exciton 
wavevector in equation (\ref{eq:exstate}) contains spin 
mixing effects between heavy and light- hole exciton states,
which forms the basis by which acoustic phonons
contribute to  spin relaxation processes and 
will be further studied in Section~\ref{sec:phonon}. 


\subsection{Exciton spin states at $\theta_{{\rm e}} = \theta_{{\rm h}} = 0$}

At $\theta_{{\rm e}} = \theta_{{\rm h}} = 0$ the electron and hole spins are
oriented in the direction of the quantization
axis and specific forms for the electron and 
hole creation operators are given by the following, firstly for heavy-hole excitons:
\begin{eqnarray}
\label{eq:opera0H}
| {\vec{k}_{{\rm e}}}, {\vec{k}_{{\rm h}}}; S \rangle_{{\rm H}}
= \sum _{\stackrel{\tau_z = 
{\pm \frac{1}{2}}   } {\sigma_{z}= {\pm \frac{3}{2}}  }}
\; 
\left ( \half \tau_z ; \,
\textstyle {3\over2} \sigma_{z} \mid S \, S_z \right )
\hat a_{\vec{k}_{{\rm e}}}^\dagger(\tau_z) \; 
\hat c_{\vec{k}_{{\rm h}},{\rm H}}^{\dagger}(\sigma_z)
 |0 \rangle
\end{eqnarray}
and, secondly, for light-hole excitons:
\begin{eqnarray}
\label{eq:opera0L}
| {\vec{k}_{{\rm e}}}, {\vec{k}_{{\rm h}}}; S \rangle_{{\rm L}}
= \sum _{\stackrel{\tau_z = 
{\pm {1\over2}}}    {\sigma_{z}= {\pm \frac{1}{2}} }}
\; \left ( \half \tau_z ; \,
 \half \sigma_{z}\mid S \, S_z \right )
\hat a_{\vec{k}_{{\rm e}}}^\dagger(\tau_z) \; 
\hat c_{\vec{k}_{{\rm h}},{\rm L}}^{\dagger}(\sigma_z)
|0 \rangle, 
\end{eqnarray}
where
$|0 \rangle$ is the electronic vacuum state of the system
representing completely filled valence band and empty conduction
bands, and the factors  $\left ( \frac{1}{2} \tau_z ; \,
\frac{3}{2} \sigma_z|S \, S_z \right )$
and $\left ( \frac{1}{2} \tau_z ; \,  
\frac{1}{2} \sigma_z |S S_z \right )$
are Clebsch-Gordon coefficients.  
By using equations (\ref{eq:opera0H}) and (\ref{eq:opera0L})
we obtain the following exciton spin states (for $S=0$ and $S=1$)
at $\theta_{{\rm e}} = \theta_{{\rm h}} = 0$:
\begin{eqnarray}
\fl
\label{eq:opera1}
| {\vec{k}_{{\rm e}}}, {\vec{k}_{{\rm h}}}; S=0 \rangle_{{\rm L}}
= 
\case1{{\sqrt2}} 
\left [a_{\vec{k}_{{\rm e}}}^\dagger(\half)\;
c_{\vec{k}_{{\rm h}},{\rm L}}^\dagger(-\half)-
a_{\vec{k}_{{\rm e}}}^\dagger(-\half)\;
c_{\vec{k}_{{\rm h}},{\rm L}}^\dagger(\half) \right]|0\rangle,
\\
\fl
\label{eq:opera2}
|{\vec{k}_{{\rm e}}}, {\vec{k}_{{\rm h}}}; S = 1 \rangle_{{\rm L}}
= \case1{{\sqrt6}} \left[a_{\vec{k}_{{\rm e}}}^\dagger(-\half)\;
c_{\vec{k}_{{\rm h}},{\rm L}}^\dagger(\half) + 
a_{\vec{k}_{{\rm e}}}^\dagger(\half)\;
c_{\vec{k}_{{\rm h}},{\rm L}}^\dagger(-\half) \right]   |0 \rangle
\nonumber 
\\
+
\case1{{\sqrt3}} 
\left[
a_{\vec{k}_{{\rm e}}}^\dagger(-\half)\;
c_{\vec{k}_{{\rm h}},{\rm L}}^\dagger(-\half)+
a_{\vec{k}_{{\rm e}}}^\dagger(\half)\;
c_{\vec{k}_{{\rm h}},{\rm L}}^\dagger(\half) 
\right] |0 \rangle,
\\
\fl
\label{eq:opera3}
| {\vec{k}_{{\rm e}}}, {\vec{k}_{{\rm h}}}; S = 1 \rangle_{{\rm H}}
= \case1{{\sqrt2}} \left [a_{\vec{k}_{{\rm e}}}^\dagger(-\half)\;
c_{\vec{k}_{{\rm h}},{\rm H}}^\dagger(\textstyle{{3\over2}})+ 
a_{\vec{k}_{{\rm e}}}^\dagger(\half)\;
c_{\vec{k}_{{\rm h}},{\rm H}}^\dagger(-\textstyle{{3\over2}})\right ] |0 \rangle.
\end{eqnarray}
The  $S = 0$ spin state of the exciton 
arises from the coupling of the electron to the 
light-hole while the $S = 1$ spin state results from the combination 
of three symmetrical spin functions of the electron and 
light-hole spin states with the spin components $S_{z} = +1, 0, -1$.
The $S = 1$ exciton can also be formed by coupling the 
components $S_{z} = \pm1$ between 
electron and heavy-hole spin states. 
Spin states $S=2$ of the heavy-hole exciton cannot couple to photons in 
the light field and are therefore known as
optically inactive dark excitons. 
Hence, the light-hole and heavy-hole excitons are characterized
by the spin projections $S_z=0,\pm1$ and $S_z=\pm1,\pm2$ of the
total angular momentum $S=1,2$ respectively.
The short range exchange interaction (given below in equation 
(\ref{eq:coulintx})) further splits the ground states of both LH and HH
excitons into doublet states, an effect also known as singlet-triplet
splitting (see \cite{Bay} for a recent review).
Hence, in low dimensional systems spin mixing effects between heavy-
and light-hole states result in a complicated energy spectrum
and the spin states for $S = 0$ or  $S = 1$ are no longer eigenfunctions of
the corresponding Hamiltonian.  
In this paper we neglect heavy-hole and light-hole mixing effects
due to exchange interactions. We also assume that the mutual conversion
between the heavy-hole and light-hole exciton states arises from the
phonon scattering processes as discussed in detail below in 
\S\ref{sec:phonon}.


\subsection{Exciton eigenvalue problem}
\label{sec:theo4}
 
The exciton eigenvalue $ E_{{\rm ex}}({\vec{K}}; S)$ 
corresponding to the eigenvector $|{\vec{K}}; S \rangle$ in
equation (\ref{eq:exstate}) is found by solving the Schr\"odinger equation:
\beq
\label{eq:schreqn0}
\left (\hat H_{{\rm kin}}+\; \hat H_{{\rm loc}}+ \;
\hat H_{{\rm int}}^{{\rm C}} +\hat H_{{\rm int}}^{{\rm exch}} 
\right)|{\vec{K}}; S\rangle
= E_{{\rm ex}}({\vec{K}}; S)|{\vec{K}} ; S \rangle,
\eeq
where the Hamiltonian is written as a sum of four terms which we now describe in turn.
$\hat H_{{\rm kin}}$ is the kinetic energy:
\begin{eqnarray}
\fl
\label{eq:schreqn1}
\hat H_{{\rm kin}} = 
\sum_{\stackrel{\vec{k}_{{\rm e}}, \vec{k}_{{\rm h}}}
{\varsigma,\eta}}\; 
\left [ E_g -
{\hbar^2 {\vec{k}_{{\rm e}}}^2 \over 2 m_{{\rm e}}} \right ] 
a_{\vec{k}_{{\rm e}}}^\dagger(\varsigma)\; a_{\vec{k}_{{\rm e}}}(\varsigma) 
\\
\nonumber
- 
\left[
{\hbar^2 {\vec{k}_{{\rm h}}}^2 \over 2 m_{{\rm L}}}\; 
c_{\vec{k}_{{\rm h}},{\rm L}}^\dagger(\eta)\; c_{\vec{k}_{{\rm h}},{\rm L}}(\eta) 
+ 
{\hbar^2 {\vec{k}_{{\rm h}}}^2 \over 2 m_{{\rm H}}} 
c_{\vec{k}_{{\rm h}},{\rm H}}^\dagger(\eta) \; c_{\vec{k}_{{\rm h}},{\rm H}}(\eta)
\right] , 
\end{eqnarray}
where $m_{{\rm e}}$ is the effective electron mass and $m_{{\rm H}}$ 
($m_{{\rm L}}$) is the heavy-hole (light-hole) mass.

$\hat H_{{\rm loc}}$ is associated with the localization 
energies of electron and hole, and is given by
\bea
\fl
\label{eq:schreqn2}
\hat H_{{\rm loc}} = {\langle p_{z_{{\rm e}}}^2 \rangle  \over 2 m_{{\rm e}} }\; 
a_{\vec{k}_{{\rm e}}}^\dagger(\varsigma)\; 
a_{\vec{k}_{{\rm e}}}(\varsigma) +
{\langle p_{z_{{\rm h}}}^2 \rangle \over 2 m_{{\rm L}} }\; 
c_{\vec{k}_{{\rm h}},{\rm L}}^\dagger(\eta)\; c_{\vec{k}_{{\rm h}},{\rm L}}(\eta) 
+ {\langle p_{z_{{\rm h}}}^2 \rangle  \over 2 m_{{\rm H}} } 
c_{\vec{k}_{{\rm h}},{\rm H}}^\dagger(\eta) \; c_{\vec{k}_{{\rm h}},{\rm H}}(\eta) ,
\eea
where 
\bea
\label{eq:well}
\langle p_{z_{{\rm e}}}^2 \rangle = 
\int_{-L_{{\rm w}} /2}^{L_{{\rm w}}/2} dz_{{\rm e}} \; \xi_{\vec{k}_{{\rm e}}}^*(z_{{\rm e}})
{\left(-\rmi\hbar\frac{\partial}{\partial z_{{\rm e}}} \right)}^2 
\xi_{\vec{k}_{{\rm e}}}(z_{{\rm e}}) , 
\eea
with a similar expression for $\langle p_{z_{{\rm h}}}^2 \rangle$,
and $L_{{\rm w}}$ denotes the thickness of the quantum well. 

$\hat H_{{\rm int}}^{{\rm C}}$ denotes the Hamiltonian due to the
Coulomb interaction which scatters an
electron in the initial state with momentum $\vec{k}$
  and spin $\varsigma$ to a final state with  momentum 
$\vec{k}+\vec{q}$ and spin $\varsigma^\prime$. Likewise  
a hole in   its initial state with momentum $\vec{k}^\prime$ 
and spin $\eta$  is scattered to a final state with
  momentum $\vec{k}^\prime-\vec{q}$ and spin $\eta^\prime$, and hence:
\begin{eqnarray}
\fl
\label{eq:coulint}
\hat H_{{\rm int}}^{{\rm C}} = 
\case12\sum_{\stackrel{\vec{k}, 
\vec{k}^\prime, \vec{q},X}{\varsigma,\eta,
\varsigma^\prime,\eta^\prime}}\; 
\left\langle {\vec{k}+\vec{q}}, \varsigma^\prime;
 \; {\vec{k}^\prime -\vec{q}},\eta^\prime
\right|U(| {\bf r_{{\rm e}}}-{\bf r_{{\rm h}}}| )  
\left|{\vec{k}^\prime},\eta; \;
{\vec{k}}, \varsigma \right\rangle \; 
\\
\nonumber
\qquad
\times
a_{\vec{k}+\vec{q}}^\dagger(\varsigma^\prime) \;
c_{\vec{k}^\prime -\vec{q},X}^\dagger(\eta^\prime)\;
c_{\vec{k}^\prime,X}(\eta) \; a_{\vec{k}}(\varsigma)  
\end{eqnarray}
where 
\[
U(|{\bf r_{{\rm e}}}-{\bf r_{{\rm h}}}|)= 
\frac{e^2}{\epsilon|{\bf r_{{\rm e}}}-{\bf r_{{\rm h}}}|},
\]
in which $\epsilon$ is the relative dielectric constant
and $X$ denotes either the heavy-hole operator H or the 
light-hole operator L. 
In the two-particle interaction matrix element
\[
\left\langle {\vec{K}+\vec{q}}, \varsigma^\prime; 
\; {\vec{K}^\prime -\vec{q}},\eta^\prime
\right|U(| {\bf r_{{\rm e}}}-{\bf r_{{\rm h}}}| ) \left|{\vec{K}^\prime},\eta; \;
{\vec{K}}, \varsigma \right\rangle,
\]
the states  
to the right of the scattering potential $U(| {\bf r_{{\rm e}}}-{\bf r_{{\rm h}}}| )$
represent the initial states while those to the left 
represent the  final scattered states.

$\hat H_{{\rm int}}^{{\rm exch}}$ denotes the Hamiltonian due to 
the exchange interaction which scatters an 
electron with spin $\varsigma$ and momentum $\vec{k}^\prime$
in the conduction band  to a hole state carrying
momentum $\vec{k}^\prime-\vec{q}$ and spin $\eta$ in  the valence
band. At the same  time,  a hole state with  momentum $\vec{k}$ 
and $\eta$ in the valence band is scattered  
 to an electron state  with momentum $\vec{k}+\vec{q}$
and spin $\varsigma$ in the conduction band: 
\begin{eqnarray}
\fl
\label{eq:coulintx}
\hat H_{{\rm int}}^{{\rm exch}} 
= \case12
\sum_{\stackrel{\vec{k}, \vec{k}^\prime, 
\vec{q}, X}{\varsigma,\eta}}\; 
\left\langle 
{\vec{k}^\prime-\vec{q}}, \eta; \; {\vec{k}+\vec{q}},\varsigma
\right|U(| {\bf r_{{\rm e}}}-{\bf r_{{\rm h}}}| )  
\left|\vec{k},\eta; \;
{\vec{k}^\prime}, \varsigma \right\rangle \; 
\\
\nonumber\times
c_{\vec{k}^\prime -\vec{q},X}^\dagger(\eta)\;
a_{\vec{k}+\vec{q}}^\dagger(\varsigma) \;
c_{\vec{k},X}(\eta) \; a_{\vec{k}^\prime}(\varsigma)  .
\end{eqnarray}

By using equations 
(\ref{eq:creop},\ref{eq:holesubband1}-\ref{eq:schreqn1},%
\ref{eq:coulint},\ref{eq:coulintx}) and by setting
$\theta_{{\rm e}} = \theta_{{\rm h}} = 0$ we obtain, after some manipulation,
the following two coupled equations for the unknown
coefficient functions $F_{{\rm H}}({\vec{k}_{{\rm e}}},{\vec{k}_{{\rm h}}};\varsigma,\eta)$ 
and $F_{{\rm L}}({\vec{k}_{{\rm e}}},{\vec{k}_{{\rm h}}};\varsigma,\eta)$
which are defined by equation~(\ref{eq:exstate}):
\begin{eqnarray}
\fl
\label{eq:coeffs1}
0=
\left [E_{{\rm e}}(\vec{k}_{{\rm e}})
+ E_{{\rm H}}(\vec{k}_{{\rm h}})-E_{{\rm ex}}^{{\rm H}}(\vec{K}) 
 \right ] F_{{\rm H}}({\vec{K},\vec{k}};\varsigma,\eta) 
\\
\nonumber
+ \sum_{\tau_z,\sigma_z} \; 
\int d{\vec{k}^\prime}
\;  \left ( V_{{\rm C}}(\vec{k},\vec{k}^\prime ; H) 
-(1-S)\; V_{{\rm exch}}(\vec{k},\vec{k}^\prime ; H) \right )
\\ 
\fl
\times \nonumber 
\left [ F_{{\rm H}}({\vec{K},\vec{k}^\prime};\varsigma,\eta)
\; d_{\varsigma \frac{1}{2} \; \tau_z}^{1/2}(\phi_{{\rm e}})
\; d_{\eta \frac{3}{2} \; \sigma_z}^{3/2}(\phi_{{\rm h}})  
+  F_{{\rm L}}({\vec{K},\vec{k}^\prime};\varsigma,\eta) \;
\; d_{\varsigma \frac{1}{2} \; \tau_z}^{1/2}(\phi_{{\rm e}})
 d_{\eta \frac{1}{2} \; \sigma_z}^{3/2}(\phi_{{\rm h}}) \right ]
\end{eqnarray}
and
\begin{eqnarray}
\label{eq:coeffs2}
\fl
0=\left [ E_{{\rm e}}(\vec{k}_{{\rm e}})+ 
E_{{\rm L}}(\vec{k}_{{\rm h}})
-E_{{\rm ex}}^{{\rm L}}(\vec{K}) 
\right ] F_{{\rm L}}({\vec{K},\vec{k}};\varsigma,\eta)
\\
\nonumber 
+ \sum_{\tau_z,\sigma_z} \; \int d{\vec{k}^\prime}
\; \left ( V_{{\rm C}}(\vec{k},\vec{k}^\prime ; L) -
(1-S) \; V_{{\rm exch}}(\vec{k},\vec{k}^\prime ; L) \right )\; 
\\
\fl
\times \nonumber 
\left [ F_{{\rm H}}({\vec{K},\vec{k}^\prime};\varsigma,\eta) 
\; d_{\varsigma \frac{1}{2} \; \tau_z}^{1/2}(\phi_{{\rm e}})
\;d_{\eta \frac{3}{2} \; \sigma_z}^{3/2}(\phi_{{\rm h}})
+  F_{{\rm L}}({\vec{K},\vec{k}^\prime};\varsigma,\eta) \;
\; d_{\varsigma \frac{1}{2} \; \tau_z}^{1/2}(\phi_{{\rm e}})
\;d_{\eta \frac{1}{2} \; \sigma_z}^{3/2}(\phi_{{\rm h}}) \right ] ,
\end{eqnarray}
where $E_{{\rm e}}(\vec{k}_{{\rm e}})$ and  $E_{{\rm H}}(\vec{k}_{{\rm h}})$ 
($E_{{\rm L}}(\vec{k}_{{\rm h}})$)
denote the in-plane energies of the electron and 
heavy-hole (light-hole)
respectively, and $E_{{\rm ex}}^{{\rm H}}(\vec{K})$ 
($E_{{\rm ex}}^{{\rm L}}(\vec{K})$) denotes
the heavy-hole (light-hole) exciton energy 
in the plane of the quantum well.
The angles $\phi_{{\rm e}}$ and $\phi_{{\rm h}}$ are defined by equations~(\ref{eq:holesubband1}) 
and (\ref{eq:holesubband}).

The integral forms of $V_{{\rm C}}(\vec{k},\vec{k}^\prime ; X)$  and
$V_{{\rm exch}} (\vec{k},\vec{k}^\prime ; X)$ which are the  
in-plane Fourier transforms of the Coulomb 
potential, with  $X = H$ ($\sigma_{z},\sigma_{z}^\prime = 
\pm \frac{3}{2}$) or $X = L$ ($\sigma_{z},\sigma_{z}^\prime = 
\pm \frac{1}{2}$), are given by:
\begin{eqnarray}
\fl
\label{kfac1}
V_{{\rm C}}(\vec{k},\vec{k}^\prime ; X) =
-\frac{e^{2}}{2\pi\epsilon }\,
 \frac{1}{|\vec{k}-\vec{k}^\prime|}
\sum _{\sigma_z,\sigma_z^\prime} \; 
\e^{\rmi\, (\sigma_z \, \phi_{{\rm h}} - \sigma_z^\prime \,\phi_{{\rm h}}^\prime)}\,
\\
\nonumber
\times
\int\!\!\!\! \int dz_{{\rm e}}\, dz_{{\rm h}}\, 
\e^{-|\vec{k}-\vec{k}^\prime|\, |z_{{\rm e}}-z_{{\rm h}}|}\, 
{\xi^*_{|{\vec{k}+\vec{q}}|}}(z_{{\rm e}}) \;
\xi_{|\vec{k}^\prime -\vec{q}|}^{* \, \sigma_z}(z_{{\rm h}})
\;
\xi_{|\vec{k}|}(z_{{\rm e}})
\;
\xi_{|\vec{k}^\prime|}^{\sigma_z^\prime}(z_{{\rm h}}),
\end{eqnarray}
 and
\begin{eqnarray}
\label{kfac2}
\fl
V_{{\rm exch}} (\vec{k},\vec{k}^\prime ; X) =
-\frac{e^{2}}{2\pi\epsilon }\, 
\frac{1}{|\vec{k}-\vec{k}^\prime|}
\sum _{\sigma_z,\sigma_z^\prime} \; 
\e^{\rmi\, (\sigma_z \, \phi_{{\rm h}} - \sigma_z^\prime \,\phi_{{\rm h}}^\prime)}\,
\\
\nonumber
\times
\int \!\!\!\!\int dz_{{\rm e}}\, dz_{{\rm h}}\, 
\e^{-|\vec{k}-\vec{k}^\prime|\, |z_{{\rm e}}-z_{{\rm h}}|}\, 
 {\xi^*_{|{\vec{k}+\vec{q}}|}}(z_{{\rm e}}) \;
\xi_{|\vec{k}^\prime -\vec{q}^\prime|}^{* \, \sigma_z}(z_{{\rm h}})
\;
\xi_{|\vec{k^\prime}|}(z_{{\rm e}}) 
\;
\xi_{|\vec{k}|}^{\sigma_z^\prime}(z_{{\rm h}}).
\end{eqnarray}

Equations (\ref{eq:coeffs1}) and (\ref{eq:coeffs2})
give rise to strong nonparabolic 
exciton centre-of-mass dispersions and 
require a large basis function set and  
sophisticated numerical techniques 
in order to be solved, even for the particular case
$\theta_{{\rm e}} = \theta_{{\rm h}} = 0$.
As a result of the nonparabolicity there is no unique definition for the 
mass of the exciton, which has resulted in  
a range of mass values in the literature \cite{win}.
The hole spin components also become mixed and some authors \cite{Une} have found it
convenient to use the parity-flip mechanism to incorporate changes in 
parity in the valence band states.

The centre-of-mass momentum wavevector ${\vec{K}}$ and 
the relative wavevector $\vec{k}$ are defined in terms of the electron
and hole wavevectors ($\vec{k}_{{\rm e}}$ and $\vec{k}_{{\rm h}}$ respectively) 
by:
\bea
\label{eq:coord}
\vec{K} & = & \vec{k}_{{\rm e}} - \vec{k}_{{\rm h}},  \nonumber \\
\vec{k} & = & (1-{\bf \vartheta})\;
 \vec{k}_{{\rm e}} + {\vartheta} \;\vec{k}_{{\rm h}},
\eea
where the parameter ${\bf \vartheta}$ is generally 
used in a scalar form \cite{oh,win,thila}
which conveniently allows the decoupling of the exciton centre-of-mass motion 
from its relative motion. In more rigorous calculations
involving the heavy-hole and light-hole dispersion relations,
${\bf \vartheta}$ has been used as a  tensor \cite{alte1,alte2}.
Recenty, Siarkos et al \cite{sia} used 
a variational approach involving minimization of  
the exciton energy and   obtained optimized  
values of ${\bf \vartheta}$
in the range $0 \leqslant {\vartheta} \leqslant  0.5$
at  $|\vec{K}| \leqslant 0.4\; {\rm nm}^{-1}$. For our work, we select
suitable values for ${\vartheta}$ 
for the heavy- and light-hole $1s$ exciton 
by extending the expression derived by
Kane \cite{kane1} for the energy shift of the 
bulk $1s$ exciton from the conduction
band minimum to a low dimensional ${\vec{K}}-$space:
\bea
\label{eq:exciKE}
E_{{\rm ex}}^{X}(\vec{K}) &=  -R_y - 5 g_1(E)\,\gamma_2^2
+ \frac{\hbar^2 {\vec{K}}^2}{2} \left ( \frac{1}{M_a} \pm
\frac{1}{M_c} \right ),  
\nonumber \\
\frac{1}{M_a} &=  \frac{1}{M_0}- \frac{40}{3} g_3(E) \,
s^2 \, \gamma_2^2  
\nonumber \\
\frac{1}{M_c} &=  \frac{2 s^2}{m_0}\, \gamma_2^2 
\eea
where the $+$ sign and $X = H$ correspond to the
heavy-hole exciton while the $-$ sign and 
$X = L$ correspond to the light-hole exciton.  
$R_y$ is the effective Rydberg constant, 
$m_0$ is the free electron mass and  
\[
M_0 = \gamma_1^{-1} + {m_{{\rm e}}}, \qquad  
s = \frac{{m_{{\rm e}}}^{-1}}
{\gamma_1+{m_{{\rm e}}}^{-1}}.
\] 
The functions $g_1(E)$ and
$g_3(E)$ given by Kane \cite{kane1} are assumed to apply
in low dimensional systems, and the constants $\gamma_1,\gamma_2,\gamma_3$ are discussed in the
Appendix. By using $E =  - R_y$ and the approximation  $\gamma_3 \approx
\gamma_2$ we obtain:
\bea
\label{eq:exciKE2}
\alpha_{{\rm e}} & = & \vartheta = \frac{m_{{\rm e}}} {\gamma_1^{-1}+ m_{{\rm e}}} +
2 s^2 \, \gamma_2 \, m_{{\rm e}}  \left [\pm \,1 - 
\frac{5 \gamma_2} {\gamma_1+ m_{{\rm e}}^{-1}}  \right ],
  \nonumber \\
\alpha_{{\rm h}} & = & 1- \vartheta. 
\eea
With the value $m_{{\rm e}} = 0.065 \; m_0$ for the effective electron mass and $\gamma_1, \gamma_2$
as given in table\ \ref{parameter} 
for the GaAs/Al$_{x}$Ga$_{1-x}$As material system, we obtain
$\alpha_{{\rm e}}$ = 0.377 for the heavy-hole exciton
and $\alpha_{{\rm e}}= 0.116$ 
for the light-hole exciton. These values are
consistent with those obtained in reference \cite{sia}. 

In order to obtain an explicit form for the exciton wavevector,
we neglect the relatively weak exchange interaction term 
$V_{{\rm exch}} (\vec{k},\vec{k}^\prime,X)$ 
involving an overlap of the conduction and valence bands. 
We now write equation~(\ref{eq:exstate}) in the form:
\beq
\label{eq:exstatex}
|{\vec{K}}; S  \rangle = | \vec{K}; S \rangle_{{\rm H}}
+ | \vec{K}; S \rangle_{{\rm L}}
\eeq
and use equations (\ref{eq:coeffs1}) and (\ref{eq:coeffs2}) to obtain
(for the heavy-hole exciton):
\begin{eqnarray}
\fl
\label{eq:exfunctH}
|\vec{K}; S \rangle_{{\rm H}}
=  
\sum _{\vec{k}_{{\rm e}}, \vec{k}_{{\rm h}}, q_z} \, 
\sum _{\stackrel{\mu = \pm1, \tau_z = 
{\pm \frac{1}{2}}}{\sigma_z= {\pm \frac{3}{2}}}}
\; \; d_{\mu  \frac{1}{2} \; \tau_z}^{1/2}(\phi)
\;d_{\mu  \frac{3}{2} \; \sigma_z}^{3/2}(\phi)  
\; \delta_{\vec{K},{\vec{k}_{{\rm e}}} - {\vec{k}_{{\rm h}}}} \; \;  
\Psi_{1s}(\alpha_{{\rm e}} {\vec{k}_{{\rm h}}} 
+ \alpha_{{\rm h}} {\vec{k}_{{\rm e}}})  
\nonumber \\
\times
F_{{\rm e}}(q_z) \, F_{{\rm h}}(q_z) \, \, 
\hat a_{\vec{k}_{{\rm e}}}^\dagger(\tau_z) \; 
\hat c_{\vec{k}_{{\rm h}}}^{\dagger}(\sigma_z )
|0\rangle,
\end{eqnarray}
and (for the light-hole exciton):
\begin{eqnarray}
\fl
\label{eq:exfunctL}
|\vec{K}; S \rangle_{{\rm L}}
= \sum _{\vec{k}_{{\rm e}}, \vec{k}_{{\rm h}}, q_z} \, 
\sum _{\stackrel{\mu = \pm1, \tau_z = {\pm \frac{1}{2}}}
{\sigma_z = {\pm \frac{1}{2}}}}
\; \; d_{\mu \frac{1}{2} \; \tau_z }^{1/2}(\phi)
\;d_{\mu  \frac{3}{2} \; \sigma_z}^{1/2}(\phi)  
\; \delta_{\vec{K},{\vec{k}_{{\rm e}}}- {\vec{k}_{{\rm h}}}} \; \;  
\Psi_{1s}(\alpha_{{\rm e}} {\vec{k}_{{\rm h}}} +
\alpha_{{\rm h}} {\vec{k}_{{\rm e}}}) 
\nonumber \\
\times
F_{{\rm e}}(q_z) \, F_{{\rm h}}(q_z) \, \, 
\hat a_{\vec{k}_{{\rm e}}}^\dagger(\tau_z) \; 
\hat c_{\vec{k}_{{\rm h}}}^{\dagger}(\sigma_z)
|0\rangle 
\end{eqnarray}
where $\phi$ is the angle of scattering 
and $\mu$ denotes  the polarization state of the exciton.
We have also dropped the $X = H (L)$ label from the hole
creation operator as  we have restricted the 
azimuthal angles to $\theta_{{\rm e}} = \theta_{{\rm h}} = 0$,
and therefore use the same notation
$\hat c_{\vec{k}}^{\dagger}({\sigma_{z}})$ for the  
heavy-hole and light-hole  creation operator.

The form factors $F_{{\rm e}}(q_z)$ and $F_{{\rm h}}(q_z)$ are given by:
\bea
\label{form}
F_{{\rm e}}(q_z) &=&  \int dz_{{\rm e}} \, \e^{\rmi q_z \, z_{{\rm e}}}\, 
{\left|\xi_{|\vec{k}_{{\rm e}}|}(z_{{\rm e}})\right|}^2,  \nonumber 
\\
F_{{\rm h}}(q_z)  &=&  \int dz_{{\rm h}} \, \e^{\rmi q_z \, z_{{\rm h}}}\, 
{\left|\xi_{|\vec{k}_{{\rm h}}|}^{\sigma_z}(z_{{\rm h}})\right|}^2.
\eea
The exciton wavevectors in equations (\ref{eq:exstatex}) -
(\ref{eq:exfunctL}) generalize the form used
in earlier work \cite{oh} in which we assumed a simple 
two-band model without taking into account the 
fourfold degeneracies of the upper valence energy bands. 
$\Psi_{1s}(\alpha_{{\rm e}} {\vec{k}_{{\rm h}}}+ \alpha_{{\rm h}} {\vec{k}_{{\rm e}}})$ 
is the wavefunction of a hydrogen type system
in momentum space, which depends on the
relative electron-hole separation. This function
determines the intrinsic properties of the exciton and
forms the basis of the fractional dimensional 
approach detailed in the next section.


\section{The fractional dimensional formalism}
\label{sec:frac}

In the fractional dimensional approach
the  confined ``exciton and quantum well", ``exciton and 
quantum wire" or ``exciton and disk" system is modelled using conventional
Hilbert spaces which carry a fractional dimensional parameter, and in which the composite
exciton system behaves in an unconfined manner, and so the
fractional dimension is related to the degree of
confinement of the physical system. Such an approach 
introduces simplicity and utility and 
has been used in the investigation of several important processes 
in low dimensional systems \cite{exscat}-\cite{expho}.
Very recent calculations in quantum wires \cite{qwire} 
with moderate strength of coupling between the subbands
have shown that the fractional dimensional approach
provides strikingly good agreement with results obtained
using  models that require intensive computational efforts. In recent work 
\cite{max1} we have developed algebraic properties of various quantum 
mechanical systems using fractional dimensions.
These studies demonstrate the useful and important role of the dimensionality 
(denoted here by $\alpha$, following \cite{qwire}) which interpolates between  
zero in ideal quantum dots to three in an exact three dimensional system, such as 
an infinitely wide quantum well. 

The binding energy of a fractional dimensional exciton \cite{lef} is given by:
\bea
\label{eq:exbind}
E_b = -{R_y \over \left (n+{\alpha-3\over2} \right )^2}\;, 
\label{excienergy}
\eea
where $n = 1,2,\ldots$ is
the principal quantum number of the exciton 
internal state  and  $R_y$ is the effective Rydberg constant.
The function $\Psi_{1s}(\vec{k})$ appearing in equations~(\ref{eq:exfunctH}) and 
(\ref{eq:exfunctL}), where 
$\vec{k}=\alpha_{{\rm e}} {\vec{k}_{{\rm h}}}+\alpha_{{\rm h}} {\vec{k}_{{\rm e}}}$,
has been obtained (Thilagam \cite{exscat}) in the form: 
\bea
\fl
\label{eq:kspace}
\Psi_{1s}(\vec{k}) = (4 \pi)^{{\alpha \over 4} - {1 \over 4}}
(\alpha-1)^{{\alpha \over 2}+ {1 \over 2}}
\sqrt{\Gamma \left [\case12 (\alpha-1) \right ] }
(a_{_B})^{\alpha \over 2} 
\left(1+ \left [\half( \alpha-1) \,|\vec{k}|\, a_{_B} \right ]^2\right)^{-{\alpha +1 \over 2}}
\eea
where $a_{_B}$ is the three-dimensional Bohr radius of the exciton.
$\Psi_{1s}(\vec{k})$ yields the expected forms in the 
exact three dimensional and two dimensional limits. 
The dimensional parameter $\alpha$ characterizes the degree of
``compression" of the confined exciton and can be calculated in
several ways (see for instance references~\cite{lef}-\cite{matosa}).
Once $\alpha$ is determined, many useful 
properties (see \cite{exscat}-\cite{expho}, \cite{lef})
can be determined by substituting its value into the 
appropriate equation that corresponds to a physical
quantity.  Recently, Escorcia et al.~\cite{don} used variationally determined
envelope functions to calculate the wavefunctions of the 
exciton relative motion in quantum wires and disks. Like earlier
approaches \cite{lef,matosa}, the technique 
of evaluating $\alpha$ in reference \cite{don}
does not include valence band coupling effects between  
the heavy- and light-hole bands. For our work, we improve on earlier methods 
\cite{lef}-\cite{matosa}
by utilizing realistic values of $\alpha$ that is
evaluated by including valence band coupling effects through a 4 $\times$ 4 
Baldereschi-Lipari Hamiltonian \cite{alte1,alte2} which describes the 
{\it relative motion} of the exciton.  This allows salient features due to the 
anisotropic exciton band model to be incorporated in $\alpha$.  

\begin{figure}
\centerline{\epsfxsize 130mm \epsfbox{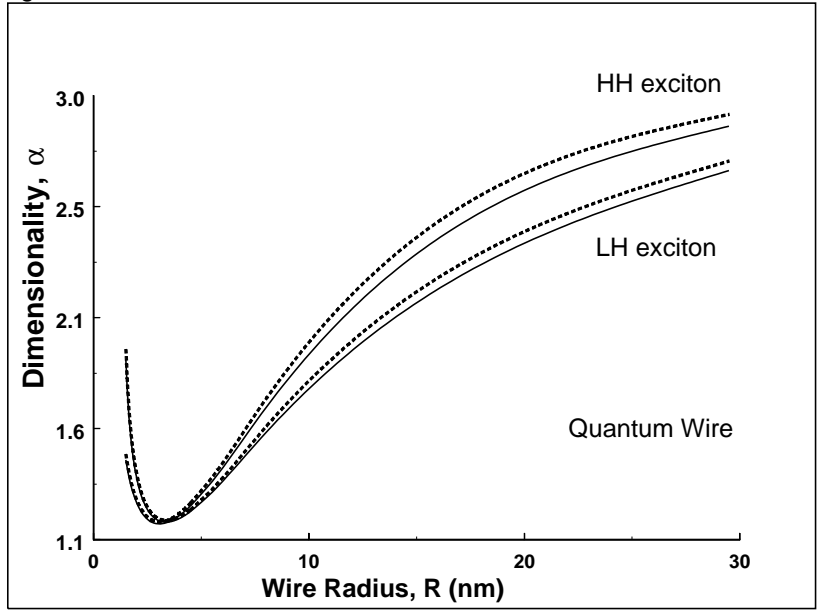}}
\caption{
\label{fig1}
Dimensionalities of the heavy-
and light-hole excitons as functions of the quantum wire
radius $R$ in GaAs/Al$_{x}$Ga$_{1-x}$As using
the parameters in table\ \ref{parameter}. The dashed
lines correspond to calculations in which off-diagonal elements
of the $4\times4$ Baldereschi-Lipari Hamiltonian
(references~\cite{alte1,alte2}) are neglected.
}
\end{figure}

We approximate the exciton ground-state wavefunction
in quantum wires and quantum disks by 
equations (\ref{eq:exstatex}-\ref{eq:exfunctL}) with appropriate
electron and hole wavefunctions 
$\xi_{|\vec{k}_{{\rm e}}|}(\bf{r_{{\rm e}}})$ and 
$\xi_{|\vec{k}_{{\rm h}}|}^{\sigma_z}(\bf{r_{{\rm h}}})$ 
determined (following references \cite{haug,bro}) 
by the quantization length $R$ which
is the radius of the quantum wire or quantum disk where the latter is 
also specified by a thickness $L$.  The function $\Psi_{1s}(\vec{k})$
is modelled using equation (\ref{eq:kspace}) with  predetermined values of $\alpha$.  
In figure~\ref{fig1} we compare the dimensionalities of the heavy-
and light-hole excitons as functions of the quantum wire 
radius $R$ in GaAs/Al$_{x}$Ga$_{1-x}$As using
the parameters listed in table\ \ref{parameter}. 
We have neglected the effective mass and dielectric mismatches
between the host materials.


At large $R$ values the dimensionality of the
light-hole exciton is noticeably smaller than that of the heavy-hole
exciton. The dimensionalities reach minimum values at the
strong confinement limit of $R_m \approx  3.5 \times 10^{-9}$ m. 
For $R \leqslant R_m$  the excitonic wavefunction spreads 
beyond the well material and leads to a rapid increase of the
dimensionalities as shown in figure~\ref{fig1}.
The effect of including the off-diagonal 
elements of the Luttinger Hamiltonian through the 
4 $\times$ 4 Baldereschi-Lipari Hamiltonian \cite{alte1,alte2}
can be seen in the difference between
the solid curves and dashed curves (in which the
off-diagonal elements of the 4 $\times$ 4 Hamiltonian 
are neglected). This figure shows that the ideal lower limit of
$\alpha=1$ is never attained. Moreover, because $\alpha \geqslant 2$ for
$R \geqslant 10^{-8}$m, dimensionalities in quantum wires are not
restricted to the range $1 \leqslant \alpha \leqslant 2$.
These results also indicate that it is not appropriate to use 
equation (\ref{eq:exbind})
to estimate $\alpha$ in quantum wires.


\section{Exciton-phonon interaction}
\label{sec:phonon}

Two main mechanisms are considered when analysing
exciton-phonon interactions in semiconductors: the
deformation potential and the piezoelectric interaction.
We write the Hamiltonian $\hat H_{{\rm ex-ph}}^{{\rm DP}} $
describing the exciton-phonon interaction, due to
the deformation potential, in terms of electron
and hole creation and annihilation operators:
\bea
\fl
\label{eq:exph}
\hat H_{{\rm ex-ph}}^{{\rm DP}}  =  
\sum_{\vec{k}, \vec{q},q_z,\tau_z^\prime,\sigma_{z}^\prime} 
\left[  M_{\tau_z}^{\lambda}({\vec{q}},q_z)\, 
\hat a_{{\vec{k} + {\vec{q}}}}^\dagger(\tau_z) \, 
 \hat a_{{\vec{k}}}(\tau_z^\prime)
-
M_{\sigma_{z}}^{\lambda}({\vec{q}},q_z) \,
\hat c_{{\vec{k} + {\vec{q}}}}^\dagger(\sigma_{z}) \,
 \hat c_{{\vec{k}}}(\sigma_{z}^\prime)
\right] 
\nonumber 
\\
\times
\left[ b^\dagger_{\lambda}({-\vec{q}},-q_z) + 
 b_{\lambda}({\vec{q}},q_z) \right],
\eea
where the phonon wave vector is given by ${\bf q} = (q_x,q_y,q_z) 
= ({\vec{q}},q_z)$, and where $b_{\lambda}^{\dagger}({\vec{q}},q_z)\,$ 
(respectively $b_{\lambda}({\vec{q}},q_z)\,$) denotes the creation (respectively annihilation) 
operator of a $\lambda-$mode phonon.  The $\lambda$-mode is
denoted ${\rm LA}$ for longitudinal acoustic phonons
and ${\rm TA}$ for transverse acoustic  phonons. 
The strain tensor $\epsilon_{ij}$ due to
the various  $\lambda-$mode phonons is 
 written in terms of normal-mode coordinates as  
\bea
\label{strain}
\fl
\epsilon_{ij} = \sum_{{\bf q},\lambda} 
\frac{\rmi}{2} \sqrt{ \frac{\hbar} 
{2\rho\, V\, \omega_{{\bf q} \, \lambda}}}
 \left [ b_{\lambda}^{\dagger}({\vec{q}},q_z) \, +  
b_{\lambda}({\vec{q}},q_z) \right ]  
(\hat{e}_{i\lambda} \, q_j +  \hat{e}_{j\lambda} \,
      q_i)\, \e^{\rmi{\bf q \cdot r}} \ ,
\eea
where $\rho$ is the  mass density of the material. 
The acoustic phonon energy spectrum is determined by
 $\omega_{_{{\bf q} {\rm LA}}}= \upsilon_{_{{\rm LA}}} |\vec{q}| $ 
for the longitudinal mode and $\omega_{_{{\bf q} {\rm TA}}} 
= \upsilon_{_{{\rm TA}}} |\vec{q}|$ for the transverse mode, with 
$\upsilon_{_{{\rm LA}}}$ and $\upsilon_{_{{\rm TA}}}$ 
denoting the corresponding sound
velocities. The term  $\hat{e}_{i\lambda}$ 
represents the unit vector of polarization
of the $\lambda$-phonon  along the $i$-direction. 
The longitudinal mode $\hat{e}_{_{{\rm LA}}}$ and the
two  transverse modes $\hat{e}_{_{{\rm TA1}}}$ and
$\hat{e}_{_{{\rm TA2}}}$ take the form \cite{hren}:
\bea
\label{eq:modes}
\hat{e}_{_{{\rm LA}}} &=&  \frac{1}{|\vec{q}|}\, 
\pmatrix{q_x \cr  q_y\cr q_z },
\nonumber  
\\
\hat{e}_{_{{\rm TA1}}} &=&  \frac{1}{|\vec{q}|} 
\, 
\pmatrix{ q_y \cr - q_x \cr  0 }
, \nonumber  
\\
\hat{e}_{_{{\rm TA2}}} &=&  \frac{1}{|{\bf{q}}|\,|\vec{q}|} \, 
\pmatrix{ q_x \,q_z \cr   q_y \, q_z \cr  - q_x^2 + q_y^2},
\eea
where the TA1-mode (TA2-mode)  corresponds to the transverse phonon
polarized along the $x$ ($y$)-direction.

The matrix elements $M_{\tau_z}^{\lambda}({\vec{q}},q_z)$ 
and $M_{\sigma_z}^{\lambda}({\vec{q}},q_z)$ in
equation (\ref{eq:exph}) depend on the  precise nature of the
exciton-phonon interactions, such as the deformation potential interaction, 
the piezoelectric interaction, and the phonon mode. We have explicitly included the spin indices
to distinguish the hole states involved in the interaction. Due to the isotropic nature
of the electron related interaction with acoustic phonons
in cubic III-V semiconductors \cite{gant}, we obtain: 
\bea
\label{eq:deforme}
M_{\pm \frac{1}{2}}^{\rm LA}({\vec{q}},q_z)
&=&  \sqrt{ \frac{\hbar \, 
|{\bf q}|} {2 \rho\, V\, \upsilon_{_{{\rm LA}}}}} \, \Xi_{\rm c} \nonumber  \\
M_{\pm \frac{1}{2}}^{\rm TA}({\vec{q}},q_z)
&=&  0,
\eea
where $\Xi_{\rm c}$ is the deformation potential constant 
associated with the conduction band. 
In using equation (\ref{eq:deforme}) some papers (for example \cite{bes})
have neglected the deformation potential coupling to the
TA phonon mode. The importance of including
interactions due to TA phonons becomes clear in the context of
hole-phonon interactions, which we consider next.

The component of the hole related
interaction in the exciton-phonon Hamiltonian is described by the
Bir-Pikus Hamiltonian $H_{{\rm BP}}$ \cite{bir1,bir2} which takes the form
\begin{equation}
\label{eq:BP}
H_{{\rm BP}} = \left( \begin{array}{cccc} 
F & H & I & 0 \\ H^* & G & 0 &
I \\ I^* & 0 & G & -H \\ 0 & I^* & -H^* & F
\end{array} \right)\ 
\end{equation}
where the matrix elements are defined (following \cite{bir1}) 
in terms of deformation potential
tensor components $\Xi_{i j}^A$ and strain 
tensor components $\epsilon_{ij}$ by:
\[
A = \sum_{i j} \, \Xi_{i j}^A \; \epsilon_{ij}, \qquad
(A = F, G, H, I )
\]
where
\bea
\label{eq:BPcoeff}
\Xi_{11}^F =  \Xi_{22}^F = a
&\Xi_{3 3}^F = b, 
\nonumber  \\
\Xi_{11}^G =  \Xi_{22}^G = \case13(a + 2 b) 
&\Xi_{3 3}^G = \case13 (4a - b),
\nonumber  \\
\Xi_{1 3}^H =  \Xi_{3 1}^H = - \rmi \, c 
&\Xi_{2 3}^H = \Xi_{3 2}^H = - \rmi \, \Xi_{1 3}^H, 
\nonumber  \\
\Xi_{1 1}^I =  -\Xi_{2 2}^I = \case1{\sqrt{3}}(a-b) 
\qquad 
&\Xi_{1 2}^I = \Xi_{2 1}^I = - \rmi \, c,
\eea
with all other components $\Xi_{i j}^A = 0$. The values
$a$, $b$ and $c$ are evaluated using the matrix elements 
$ \langle X |\hat E_{x x}|X \rangle $, 
$\langle X |\hat E_{y y}|X \rangle $ and
$\langle X |\hat E_{x y}|Y \rangle  +
\langle X | \hat E_{y x}| Y \rangle$ where the tensor
operator $\hat E_{i j}$ is obtained \cite{bir2} by subtracting 
elements of the Hamiltonian of the unstrained crystal
from those in a strained crystal. 

The effective deformation potential interaction
can be obtained separately for ${\rm TA}$ 
and ${\rm LA}$ hole-phonon interactions via a
suitable transformation of the  deformation potential
tensor matrix:
\bea
\label{eq:T1}
\widetilde{\Xi}_{i^\prime  j^\prime }^A = 
 \sum_{i j} \,U_{i^\prime i} \,
U_{j^\prime j} \, \Xi_{i j}^A 
\eea
where the matrix $U$ is given by:
\bea
\label{eq:T2}
U = 
\pmatrix{
\overline q_z \,  \cos\theta_{_\parallel} & \overline q_z \,  \sin\theta_{_\parallel} & 
\overline q_{_\parallel}  \cr
-\sin\theta_{_\parallel}  & \cos\theta_{_\parallel}  & 0  \cr
\overline q_{_\parallel} \,  \cos\theta_{_\parallel}
& \overline q_{_\parallel} \,  \sin\theta_{_\parallel}  & \overline q_z
} 
\eea
where $\overline q_z = \frac{q_z}{|{\bf q}|}$,
$\overline q_{_\parallel}  = \frac{q_{_\parallel}}{|{\bf q}|}$
and  $\vec{q} = (q_{_\parallel},\theta_{_\parallel})$. 
 
By using equations (\ref{eq:BPcoeff}), (\ref{eq:T1})
and (\ref{eq:T2}) we obtain expressions
for the  matrix elements $M_{\sigma_z}^{\lambda}({\vec{q}},q_z)$
for different phonon modes $\lambda$:
\bea
\label{eq:deformh}
M_{\pm \frac{1}{2} (\pm \frac{3}{2})}^{\rm LA}({\vec{q}},q_z)
&=&  \sqrt{ \frac{\hbar \, 
|{\bf q}|} {2 \rho\, V\, \upsilon_{_{{\rm LA}}}}} \, \left [a^\prime \pm \,
\Lambda_1(\overline q_{_\parallel}, 
\overline q_z, \theta_{_\parallel}) \right ]
\nonumber  \\
M_{\pm \frac{1}{2} (\pm \frac{3}{2})}^{\rm TA1}({\vec{q}},q_z)
&=&  \sqrt{ \frac{\hbar \, 
|{\bf q}|} {2 \rho\, V\, \upsilon_{_{{\rm TA1}}}}} \, \left [b^\prime \, 
\overline q_{_\parallel} \, \overline q_z
\pm \, \Lambda_2(\overline q_{_\parallel}, 
\overline q_z, \theta_{_\parallel}) \right ]
\nonumber  \\
M_{\pm \frac{1}{2} (\pm \frac{3}{2})}^{\rm TA2}({\vec{q}},q_z)
&=&  \pm \sqrt{ \frac{\hbar \, |{\bf q}|} {2 \rho\,
V\, \upsilon_{_{{\rm TA2}}}}} \, \Lambda_3(\overline q_{_\parallel}, 
\overline q_z, \theta_{_\parallel}) 
\eea
where $a^\prime$ and  $b^\prime$ are obtained using
$a$, $b$ and $c$ (see equation (\ref{eq:BPcoeff})) and
the $+$ and $-$ signs on the right hand side of equation (\ref{eq:deformh})
 correspond to the light- and heavy-holes
respectively. $\Lambda_i \,$ ($i = 1,2,3$) are 
explicit functions of   $\overline q_{_\parallel}$, 
$\overline q_z$ and $\theta_{_\parallel}$, and $\upsilon_{_{{\rm TA1}}}$
and $\upsilon_{_{{\rm TA2}}}$ are the  respective sound 
velocities corresponding to polarization 
in the $x$ and $y$-direction.

The Hamiltonian $\hat H_{{\rm ex-ph}}^{{\rm Piez}}$
describing the exciton-acoustic phonon scattering due to the
piezoelectric effect for LA and TA phonons
can be obtained using the Bir-Pikus Hamiltonian
in equation (\ref{eq:BP}) and the matrix $U$ 
in equation (\ref{eq:T2}), in the form:
\bea
\fl
\label{eq:exphpiezo}
\hat H_{{\rm ex-ph}}^{{\rm Piez}}  =  
\sum_{\vec{k}, \vec{q},q_z,\tau_z^\prime,\sigma_{z}^\prime} 
  M_{{\rm P}}^{\lambda}({\vec{q}},q_z)\, 
\left [ \hat a_{{\vec{k} + {\vec{q}}}}^\dagger(\tau_z) \, 
 \hat a_{{\vec{k}}}(\tau_z^\prime) -
\hat c_{{\vec{k} + {\vec{q}}}}^\dagger(\sigma_{z}) \,
 \hat c_{{\vec{k}}}(\sigma_{z}^\prime) \right ]
\\\nonumber
\times
\left[ b^\dagger_{\lambda}({-\vec{q}},-q_z) + 
 b_{\lambda}({\vec{q}},q_z) \right],
\eea
where the matrix elements $ M_{{\rm P}}^{\lambda}({\vec{q}},q_z)$ 
are given by:
\bea
\label{eq:piezo}
M_{{\rm P}}^{\rm LA}({\vec{q}},q_z)
&=& 
\rmi \,\sqrt{ \frac{\hbar } {2 \rho\,\epsilon^2\,
V\, \upsilon_{_{{\rm LA}}} |{\bf q}|}} \,
\left(
3 \, e \, h_{14} \, \overline q_{_\parallel}^2
\, \overline q_z \, 
\sin 2\theta_{_\parallel}
\right)
 \, 
\nonumber  \\
M_{{\rm P}}^{\rm TA1}({\vec{q}},q_z)
&=& 
\rmi \,\sqrt{ \frac{\hbar } {2 \rho\,\epsilon^2\,
V \,\upsilon_{_{{\rm TA1}}} |{\bf q}|}} \, \left (
3 \, e \, h_{14} \, \overline q_{_\parallel}
\, ({\overline q_z}^2 - {\overline q_{_\parallel}}^2)
 \, \sin 2\theta_{_\parallel} \right )
 \, \nonumber  \\
M_{{\rm P}}^{\rm TA2}({\vec{q}},q_z)
&=& 
\rmi \,\sqrt{ \frac{\hbar } {2 \rho\,\epsilon^2\,
V \,\upsilon_{_{{\rm TA2}}} |{\bf q}|}} \, \left(
2 \, e \, h_{14} \, \overline q_{_\parallel}
\, \overline q_z \, 
(\cos 2\theta_{_\parallel} - 1) \right)
\eea
where $h_{14}$ is the piezoelectric coupling constant.
Unlike the matrix elements for 
the deformation potential $M_{\sigma_z}^{\lambda}({\vec{q}},q_z)$, which 
depend on the hole spin, the elements 
$ M_{{\rm P}}^{\lambda}({\vec{q}},q_z)$ corresponding to the piezoelectric
interaction are independent of the hole spin.


\section{Exciton decay and scattering involving acoustic phonons}
\label{sec:matrix}

In this section we derive rates of spin transitions due,
firstly, to the decay of heavy- and light-hole excitons,
secondly,  to intraband scattering
(heavy-hole to heavy-hole, light-hole to light-hole) and,
thirdly, to
interband scattering (heavy-hole to light-hole, 
light-hole to heavy-hole) due to interactions with LA and TA 
phonons.  For  the  {\it decay} process, we suppose
that initially an exciton with
wavevector ${\vec{K}}$  and spin $S$ decays into an electron hole
pair  by absorbing  an acoustic phonon. We assume
that the  wavefunction of the final  state  
can be written in terms of a free electron-hole pair as:
\bea
\label{eq:decay}
|{\vec{K}}, S \rangle_{_{{\rm decay}}} = \sum_{\vec{k},q_z,\lambda}
\;\delta_{{\vec{k}-\vec{K}},\vec{q}} \, \;
\hat a_{\vec{k}}^{\dagger}({\tau_z})
\; \hat c_{\vec{k}}^{\dagger}({\sigma_{z}}) \;
b_{\lambda}(\vec{q},q_z)\; |0\rangle \otimes |n_{\vec{q},q_z}\rangle,
\eea
where we have included the allowed $\lambda$-modes within the summation,
although the contribution of each $\lambda$-mode phonon
interaction to the decay process can be analysed
separately by excluding $\lambda$ from the summation,
and where
$n_{\vec{q},q_z}$ denotes the occupation number of phonons. 

For the  {\it scattering} process, we suppose that an exciton
is scattered  from  its initial state of wavevector ${\vec{K}}$ 
and spin $S$ to a final state with wavevector ${\vec{K^\prime}}$ 
($\neq\vec{K}$) and spin $S^\prime$ through
the following channels which may occur with or without a change of 
the exciton spin:

\begin{enumerate}
\item  
$ |{\vec{K}}; S  \rangle \quad \longrightarrow   \quad
|{\vec{K^\prime}}; S^\prime  \rangle$ with energy
$\hbar \omega_{{\vec{K}-\vec{K^\prime}, q_z}\,\lambda}$ (acoustic phonon emission),

\item  
$ |{\vec{K}}; S  \rangle \quad \longrightarrow   \quad
|{\vec{K^\prime}}; S^\prime  \rangle$ with energy
$\hbar\omega_{{\vec{K^\prime}-\vec{K}, q_z}\,\lambda}$ 
(acoustic phonon absorption).
\end{enumerate}
The spin relaxation time $\tau_{{\rm sp}}$ is calculated  using
the Fermi golden rule:
\begin{eqnarray}
\fl
\label{eq:fermi}
\frac{1}{\tau_{{\rm sp}}} 
 =  {2 \pi \over \hbar}\, \sum _{\vec{k}, \vec{q}, q_z} \, 
\sum _{\mu,\mu^\prime} \, {| \langle f| H_{{\rm int}}|i \rangle|}^2
\, \left[(f_{\vec{k}+ \vec{q}}+1)\, f_{\vec{k}} \,N_q^\lambda \, 
\delta(E_{{\rm ex}}({\vec{k}} + {\vec{q}}) -
 E_{{\rm ex}}({\vec{k}}) - \hbar \omega_{q \lambda})\right. 
\nonumber \\
+ 
\left.
(f_{\vec{k} - \vec{q}}+1)\,
f_{\vec{k}} \,(N_q^\lambda +1) \, \delta(E({\vec{k}}- {\vec{q}}) -
 E_{{\rm ex}}({\vec{k}}) + \hbar \omega_{q \lambda}) 
\right],
\eea
where  $|i \rangle$  and $|f \rangle$ denote the initial
and final states, $H_{{\rm int}}$  is the interaction operator
and $ N_q^\lambda$  is the thermalised average number of phonons
at the lattice temperature $T_{\rm lat}$, and is 
given by the Bose-Einstein distribution 
\[
N_q^\lambda=\left[\exp\left(
{h\,\omega_{q\lambda}\over k_{\rm B}T_{\rm lat}} \right)-1  \right]^{-1}.
\]
The exciton distribution function $f_{\vec{k}}$ is given by \cite{iva}:
\bea
\label{eq:iva}
f_{\vec{k}}^{-1} & = & \exp \left ( \frac{E_{{\rm ex}}({\vec{k}})- \mu_c}
{k_B T_{{\rm ex}}} \right ) - 1 \nonumber \\
 \mu_c & = & k_B T_{{\rm ex}}\, {\rm ln}\left [ 1 - \exp \left (
 \frac{-2\pi\hbar^2  n_{{\rm ex}}\alpha_{{\rm e}}  }{g\, m_{{\rm e}}\, k_{{\rm B}}  T_{{\rm ex}}} \right ) \right ]
\eea
where $T_{{\rm ex}}$ is the exciton temperature, $k_{{\rm B}}$ 
is the Boltzmann constant, $\mu_c$ is  the chemical
potential, $n_{{\rm ex}}$ is the exciton density,
$\alpha_{{\rm e}}$ is defined in equation~(\ref{eq:exciKE2}), $g$ is the spin
degeneracy factor \cite{iva} and, as before, $m_{{\rm e}}$ is the effective electron mass.
The summation in equation (\ref{eq:fermi}) includes all the initial and
final spin states of the coupled electron-hole system
and thus the relaxation time $\tau_{{\rm sp}}$ incorporates 
the effect of changes from spin-up (spin-down) to spin-down (spin-up) 
electron and hole states of the exciton.

By using the initial state given by equation (\ref{eq:exstatex}), 
the final state given by equation (\ref{eq:decay}),
and the Hamiltonian operator of exciton-phonon 
interactions due to the deformation potential in 
equation (\ref{eq:exph}), we obtain the transition matrix elements
for the {\it decay} of a heavy-hole exciton into a free electron heavy-hole pair as:
\begin{eqnarray}
\fl
\label{eq:matrixDP}
\langle f| \hat H_{{\rm ex-ph}}^{{\rm DP}}|i \rangle_{_{{\rm decay}}}^{{\rm HH}}
\\
\nonumber
\fl
= 
\left[  M_{\tau_z}^{\lambda}({\vec{q}},q_z)\,F_{{\rm e}}(q_z) \,
 \Psi_{1s}(\alpha_{{\rm h}} {\vec{K}} + {\vec{k}_{{\rm h}}})
-  M_{\pm \frac{3}{2}}^{\lambda}({\vec{q}},q_z) \, F_{{\rm h}}(q_z) \,
\Psi_{1s}({\vec{k}_{{\rm e}}} - \alpha_{{\rm e}} {\vec{K}})
 \right ] 
\\
\nonumber \times   
d_{\mu \frac{1}{2} \; \mu^\prime \frac{1}{2} }^{1/2}(\phi)
 \;  d_{\mu \frac{3}{2} \; \mu^\prime \frac{3}{2}}^{3/2}(\phi)
\;  \delta_{\vec{K},{\vec{k}_{{\rm e}}}- {\vec{k}_{{\rm h}}} \pm {\vec{q}}} \;  
\sqrt{(f_{\vec{k} \mp \vec{q}}+1)\, f_{\vec{k}}(N_q + \delta_{\pm 1,1})}
\end{eqnarray}
where the signs $+$ and $-$ in $\delta_{\pm 1,1}$ 
correspond to a phonon emission 
and absorption process respectively and vice versa in 
$f_{\vec{k} \mp \vec{q}}\,$.  
An expression similar to that
in equation (\ref{eq:matrixDP}) applies for the {\it decay}
of the light-hole exciton into a free electron- light-hole
pair when the reduced rotation matrix
$d_{\mu \frac{3}{2} \; \mu^\prime \frac{3}{2}}^{3/2}(\phi)$ 
is replaced by $d_{\mu \frac{1}{2} \; \mu^\prime 
\frac{1}{2}}^{3/2}(\phi)$ and when the matrix element
$M_{\pm \frac{3}{2}}^{\lambda}({\vec{q}},q_z)$ 
is replaced by $M_{\pm \frac{1}{2}}^{\lambda}({\vec{q}},q_z)$.
Appropriate changes should also be made
to $\alpha_{{\rm e}}$ and $\alpha_{{\rm h}}$ (which  
corresponds to the light-hole in equation (\ref{eq:exciKE2})) 
in  the exciton wavefunction 
$\Psi_{1s}(\alpha_{{\rm e}} {\vec{k}_{{\rm h}}}+ \alpha_{{\rm h}} {\vec{k}_{{\rm e}}})$.

By using equation (\ref{eq:exphpiezo}), 
we obtain the transition matrix elements for the decay of
the heavy-hole exciton  due  to piezoelectric effect  as:
\begin{eqnarray}
\fl
\nonumber
 \langle f| \hat H_{{\rm ex-ph}}^{{\rm Piez}} |i \rangle_{_{{\rm decay}}}^{{\rm HH}}
=M_{{\rm P}}^{\lambda}({\vec{q}},q_z) \, \;
\left[   \Psi_{1s}(\alpha_{{\rm h}} {\vec{K}} + {\vec{k}_{{\rm h}}})\, F_{{\rm e}}(q_z)
-  \Psi_{1s}({\vec{k}_{{\rm e}}} - \alpha_{{\rm e}} {\vec{K}})\, F_{{\rm h}}(q_z)
 \right ] 
\\
\nonumber 
 \times d_{\mu \frac{1}{2} \; \mu^\prime \frac{1}{2} }^{1/2}(\phi)
 \,  d_{\mu \frac{3}{2} \; \mu^\prime \frac{3}{2}}^{1/2}(\phi)  
  \delta_{\vec{K},{\vec{k}_{{\rm e}}}- {\vec{k}_{{\rm h}}} \pm {\vec{q}}} \;  
\sqrt{(f_{\vec{k} \mp \vec{q}}+1)\, f_{\vec{k}}(N_q + \delta_{\pm 1,1})}.
\end{eqnarray}
An expression similar to this applies for the decay
of the light-hole exciton when the reduced rotation matrix
$d_{\mu \frac{3}{2} \; \mu^\prime \frac{3}{2}}^{3/2}(\phi)$ 
is replaced by $d_{\mu \frac{1}{2} \; \mu^\prime 
\frac{1}{2}}^{3/2}(\phi)$.

The transition matrix elements corresponding
to the {\it scattering} of a {\it heavy-hole} exciton to
another {\it heavy-hole}  state due to 
deformation potential interaction is obtained as: 
\begin{eqnarray}
\fl
\label{eq:matrixDPS}
\langle f| \hat H_{{\rm ex-ph}}^{{\rm DP}}|i \rangle_{{\rm scatt-DP}}^{{\rm HH}} 
= 
\sum_{\vec{k},q_z^\prime} \left[  M_{\tau_z}^{\lambda}({\vec{q}},q_z)\,
 \, F_{{\rm e}}(q_z) \, F_{{\rm e}}(q_z^\prime) \,  \Psi_{1s}(\alpha_{{\rm h}} {\vec{K}} - {\vec{k}})
\, \Psi_{1s}(\alpha_{{\rm h}} {\vec{K^\prime}} - {\vec{k}})\right. 
\nonumber 
\\ 
\left. -   M_{\pm \frac{3}{2}}^{\lambda}({\vec{q}},q_z)
 \,F_{{\rm h}}(q_z) \, F_{{\rm h}}(q_z^\prime) \, 
\Psi_{1s}( \alpha_{{\rm h}} {\vec{K^\prime}}+{\vec{k}} \pm {\vec{q}}) \,
\Psi_{1s}( \alpha_{{\rm h}} {\vec{K}}+{\vec{k}} )
 \right ] \nonumber 
\\
\times  d_{\mu \frac{1}{2} \; \mu \frac{1}{2}}^{1/2}(\phi)\,
  d_{\mu \frac{3}{2} \; \mu \frac{3}{2} }^{3/2}(\phi)
 \delta_{\vec{K},\vec{K^\prime} \pm {\vec{q}}} \;  
\sqrt{(f_{\vec{k} \mp \vec{q}}+1)\, f_{\vec{k}}(N_q + \delta_{\pm 1,1})}
\end{eqnarray}
where the reduced rotation matrix
$d_{\mu \frac{3}{2} \; \mu^\prime \frac{3}{2}}^{3/2}(\phi)$ 
is replaced by $d_{\mu \frac{1}{2} \; \mu^\prime \frac{1}{2}}^{3/2}(\phi)$
and the matrix element $M_{\pm \frac{3}{2}}^{\lambda}({\vec{q}},q_z)$
replaced by $M_{\pm \frac{1}{2}}^{\lambda}({\vec{q}},q_z)$
for  the case of  scattering of a {\it light-hole} exciton to
another {\it light-hole}  state. For the  scattering of a {\it heavy-hole} exciton to
a {\it light-hole}  state and vice-versa, 
$d_{\mu \frac{3}{2} \; \mu^\prime \frac{3}{2}}^{3/2}(\phi)$
should be replaced  by 
$d_{\mu \frac{3}{2} \; \mu^\prime \frac{1}{2}}^{3/2}(\phi)$
instead. The transition matrix elements corresponding
to piezoelectric effect are not shown here but can be 
derived in similar manner as equation (\ref{eq:matrixDPS}).

The spin relaxation rates for various phonon mediated processes
(such as those due to the deformation potential and piezoelectric interactions), 
are finally calculated using equations (\ref{eq:matrixDP})--(\ref{eq:matrixDPS}) 
in (\ref{eq:fermi}) and by using the following 
summation over spin states (see Brink \cite{brink}, page 24):
 \bea
\label{eq:spinmatrix}
\sum _{\mu,\mu^\prime = \pm1} 
\left|d_{\mu \frac{1}{2} \; \mu^\prime \frac{1}{2}}^{1/2}(\phi)\right|^2
&=& 2 \, \nonumber  
\\
\sum _{\mu,\mu^\prime = \pm1} 
\left|d_{\mu \frac{3}{2} \; \mu^\prime  \frac{3}{2}}^{3/2}(\phi)\right|^2
&=& \case12 (1+ 3 \cos^2\phi) \, \nonumber  
\\
 \sum _{\mu,\mu^\prime = \pm1} 
\left|d_{\mu \frac{3}{2} \; \mu^\prime  \frac{1}{2}}^{3/2}(\phi)\right|^2
&=& \case32 \sin^2\phi \,.
\eea


\section{Numerical Results and Discussion}
\label{sec:discuss}

In figure~\ref{fig2}  we plot the spin relaxation time
$\tau_{{\rm sp}}$ of the heavy- and light-hole excitons as functions of
the exciton wavevector $|{\vec{K}}|$ in
GaAs/Al$_{0.3}$Ga$_{0.7}$As quantum  wells of well width 
$70\,$\AA\ and lattice temperature $T_{\rm lat}=10$ K.
 We calculate $\tau_{{\rm sp}}$ corresponding to
 intraband scatterings of heavy-hole to heavy-hole (labelled
HH $\rightarrow$ HH) due to the deformation potential
interaction using equations (\ref{eq:exph}), (\ref{eq:deforme}),
(\ref{eq:deformh}), (\ref{eq:fermi}), (\ref{eq:iva})
 and (\ref{eq:matrixDPS})
and integrating over both the in-plane 
($\vec{q}$) and perpendicular ($q_z$)
components of the phonon wavevector.
A similar approach is used to calculate $\tau_{{\rm sp}}$ due to
light-hole to light-hole scattering (LH $\rightarrow$ LH) 
via the deformation potential interaction. $\tau_{{\rm sp}}$ due to
the decay of the heavy-hole exciton (HH $\rightarrow$ e + h)
and decay of light-hole exciton (LH $\rightarrow$ e + h)
is likewise calculated using equations (\ref{eq:exph}), 
(\ref{eq:deforme}), (\ref{eq:deformh}), 
(\ref{eq:fermi}) and (\ref{eq:iva}) and the appropriate 
transition matrix elements.  The material parameters 
used in the calculation (following
\cite{iva},\cite{bir1}-\cite{Land})
are listed in table\ \ref{parameter}. 
We employ the BenDaniel-Duke model \cite{duke}
and the approach outlined in the Appendix to calculate
the electron and hole wavefunctions $\xi_{|\vec{k}_{{\rm e}}|}(z_{{\rm e}})$
and $\xi_{|\vec{k}_{{\rm h}}|}^{\sigma_z}(z_{{\rm h}})$ respectively.
We have assumed $\upsilon_{_{{\rm TA1}}} \approx
\upsilon_{_{{\rm TA2}}}$ and have neglected the 
effective mass and dielectric mismatches
between the well and barrier materials in our calculations.


\begin{figure}
\centerline{\epsfxsize 130mm \epsfbox{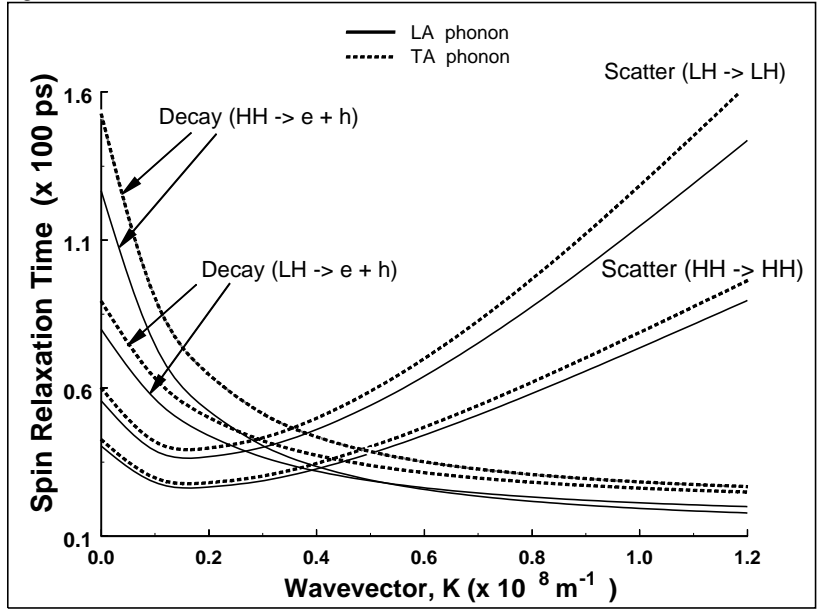}}
\caption{
\label{fig2}
The spin transition time $\tau_{{\rm sp}}$
as a function of the exciton wavevector ${\vec{K}}$ in
GaAs/Al$_{0.3}$Ga$_{0.7}$As quantum  wells. $\tau_{{\rm sp}}$
involves intraband scatterings of heavy-hole to heavy-hole (labelled as
HH $\rightarrow$ HH) and light-hole to light-hole
scattering (LH $\rightarrow$ LH), and
decay of the heavy-hole exciton (HH $\rightarrow$ e + h)
and light-hole exciton (LH $\rightarrow$ e + h).
The temperatures are $T_{\rm lat}=10$~K and $T_{\rm ex}=20$~K.
}
\end{figure}


Our results show that as the wavevector ${\vec{K}}$ increases,
the exciton spin relaxation is likely to occur
via the decay channel rather than by the scattering mechanism. Depending
on ${\vec{K}}$, the heavy-hole and light-hole excitons 
 assume different roles under
the scattering and decay processes. The light-hole exciton
is more likely to decay (at large ${\vec{K}}$) while the
heavy-hole exciton is more likely to be scattered 
(at small ${\vec{K}}$) during spin relaxation. 
It is interesting to note that the estimated times 
for $\tau_{{\rm sp}}$ are reasonably close 
for the decay and scattering mechanisms in the range
$3\times 10^7$ m$^{-1} \leqslant {\vec{K}} \leqslant 5 \times 10^7 $m$^{-1}$.
This indicates the delicate balance between decay and scattering
processes which gives rise to spin relaxation mechanisms
and the need for a careful examination of the effect of the
exciton temperature $T_{{\rm ex}}$ and the
exciton density $n_{{\rm ex}}$ on exciton
spin dynamics in low dimensional systems.

While LA and TA phonons are both effective in 
bringing about spin relaxation, the contribution
from TA phonons appears to dominate the spin dynamics of excitons
for a wide range of ${\vec{K}}$. 
The spin relaxation times (via the scattering and decay
processes) are calculated to be of the order of 10 to 100 ps which is
consistent with experimental results \cite{da1,da2} measuring 
$\tau_{{\rm sp}}$ in the order of a 100 picoseconds. 
In table\ \ref{spinval1} we have compared 
$\tau_{{\rm sp}}$(ps) corresponding to  various scattering processes in
GaAs/Al$_{0.3}$Ga$_{0.7}$As quantum wells at well width $100\,$ \AA,
${\vec{K}} = 2\times 10^7$  m$^{-1}$ and $T_{\rm lat} = 4.2$ K.
 The calculated values show that spin relaxation is dominated
by TA phonons coupled via the piezoelectric interaction
to excitons undergoing intraband (HH $\rightleftharpoons$ LH) scattering.  


\begin{figure}
\begin{center}
{\epsfxsize 130mm \epsfbox{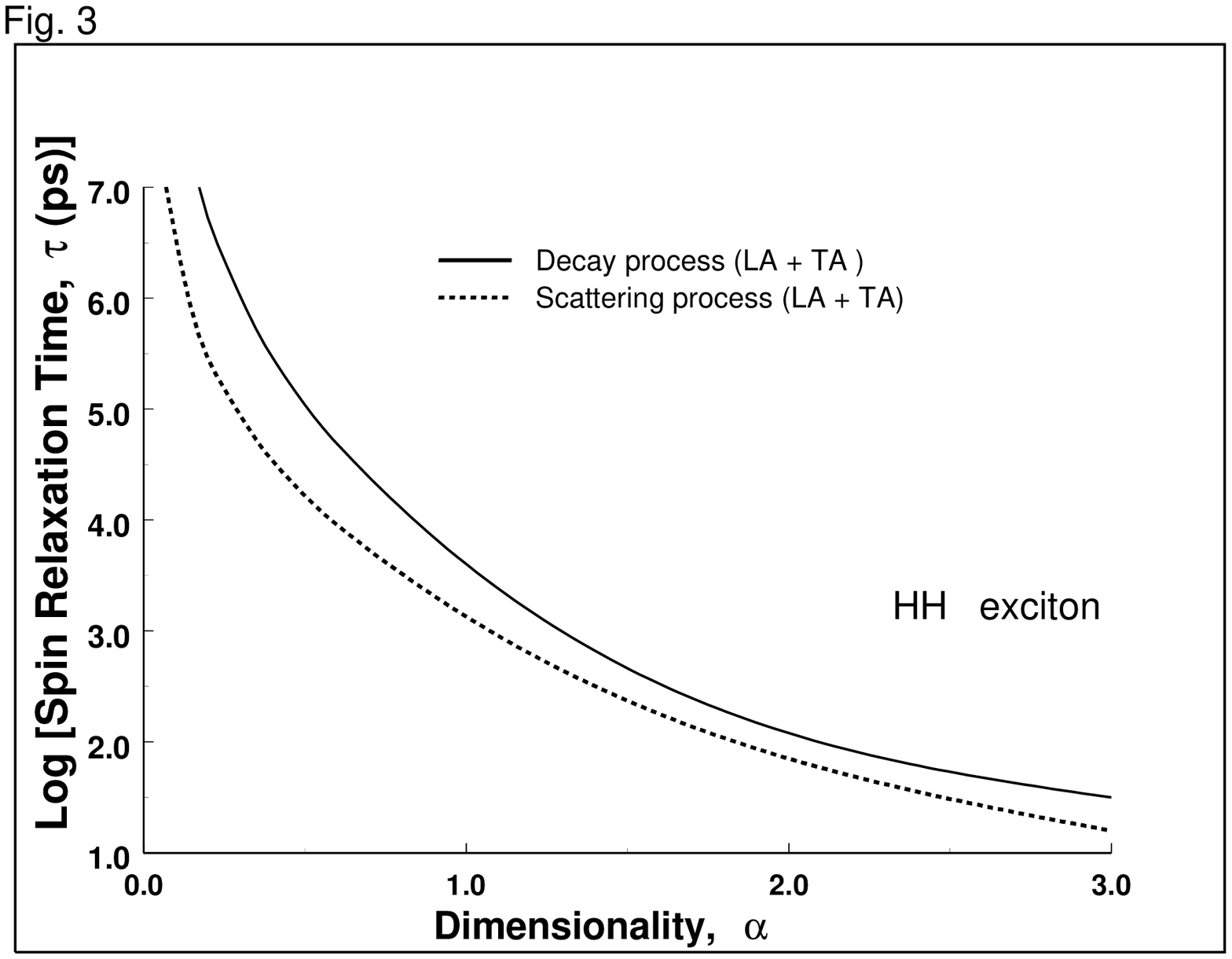}}
\end{center}
\caption{
\label{fig3}
Spin transition time $\tau_{{\rm sp}}$
as a function of the dimensionality $\alpha$ of the heavy-hole
exciton in the GaAs/Al$_{0.3}$Ga$_{0.7}$As material system
at ${\vec{K}}$ = 1 $\times 10^7  $m$^{-1}$ and $T_{\rm lat} = 4.2$~K,
$T_{\rm ex}=20$~K.
All other parameters used are listed in table\ \ref{parameter}.
The contribution of both LA and TA phonons are included
in the decay and intraband scattering calculations.
}
\end{figure}


In figure~\ref{fig3} we plot the spin transition time $\tau_{{\rm sp}}$
as a function of dimensionality $\alpha$ of the heavy-hole 
exciton in GaAs/Al$_{0.3}$Ga$_{0.7}$As material systems.
The contribution of both LA and TA phonons is included 
in  the decay and intraband scattering calculations  in which
only the deformation potential interaction is taken into 
account. Figure~\ref{fig3} shows clearly the 
 explicit dependence of $\tau_{{\rm sp}}$
on $\alpha$.  It is interesting to note that $\tau_{{\rm sp}}$
rises  as high as $1\times 10^3$ ps -- $1\times 10^6$ ps  
for $ \alpha \leqslant 1.3$. Experimental
results \cite{jap1} have shown that the spin relaxation time of excitons
in quantum disks is much higher ($\approx 1000$ ps)
than in quantum wells.
Hence at small dimensionalities there is a substantial 
increase in $\tau_{{\rm sp}}$. In table\ \ref{spinval2} we have compared 
$\tau_{{\rm sp}}$ due to the deformation potential interaction for 
different configurations of GaAs/Al$_{0.3}$Ga$_{0.7}$As systems. 
The results show the striking effect on $\alpha$ and $\tau_{{\rm sp}}$
of changing the radius of both the quantum wire and the quantum disk and its thickness.

 
\section{Conclusions}
\label{sec:concl}

We have performed  a comprehensive investigation
of exciton spin relaxation
involving LA and TA acoustic phonons for various configurations
of the GaAs/Al$_{0.3}$Ga$_{0.7}$As material system. Our results show
the explicit dependence of $\tau_{{\rm sp}}$ on $\alpha$ 
which is an important result as the relative motion of 
an exciton is  possibly best described by an intermediate dimension 
in quantum wells, wires or disks. Our results also
highlight the specific role played by TA phonons
coupled via the piezoelectric interaction to excitons. Our calculations show good agreement with 
experimental measurements and can be easily extended to  
coupled double quantum wells as well as to indirect
excitons and impurities confined in low dimensional
systems. Our results show that depending on the exciton
centre-of-mass wavevector, the 
balance between decay and scattering mechanisms resulting
in exciton spin relaxation can be delicate. 
Future developments in this field of research  will therefore
require a careful examination of the conditions 
(for example the exciton temperature, density and configuration) before an
accurate picture can be drawn of the spin dynamics 
of excitons in semiconductor systems.


\begin{table}
\caption{
\label{parameter}
Material parameters.}
\begin{indented}
\item[]
\begin{tabular}{@{}llllllll}
\br
$\rho$& $5.318\times10^3$ kg/m$^3$  && $\epsilon$ &12.9&&$\gamma_1$&6.85
\\
$\upsilon_{_{\rm TA}}$  &$1.86\times10^3$\ m/s  && $a$ &$0.31$ eV&&$\gamma_2$&2.1
\\
$\upsilon_{_{{\rm LA}}}$  &$4.25\times10^3$\ m/s  && $b$ &$2.86$ eV&&$\gamma_3$&2.9
\\
$h_{14}$ &$0.16$\ C/m$^2$ &&  $c$ &$-2.27$ eV &&  $g$&4
\\
$\Xi_c$&$-8.0 \,$ eV &&  $T_{{\rm ex}}$ &20 K &&  $n_{{\rm ex}}$&$5\times10^9$ cm$^{-2}$
\\ 
\br
\end{tabular}
\end{indented}
\end{table}


\begin{table}
\caption{
\label{spinval1}
Comparison of $\tau_{{\rm sp}}$(ps) corresponding to various 
scattering  processes  in
GaAs/Al$_{0.3}$Ga$_{0.7}$As quantum well at well width $100\,$ \AA,
$|{\vec{K}}|$ = 2 $\times 10^7 $m$^{-1}$,  with $T_{\rm lat} = 4.2$~K,
$T_{\rm ex}=20$~K.
}
\begin{indented}
\item[]
\begin{tabular}{@{}lllll}
\br
&Deform&Deform&Piezo&Piezo \\ 
&(LA)&(TA) &(LA)&(TA) 
\\
\mr
HH $\rightarrow$ HH&23&26&35&44\\ 
LH $\rightarrow$ LH&34&38&49&58\\ 
HH $\rightleftharpoons$ LH&48&63&70&79\\ 
\br
\end{tabular}
\end{indented}
\end{table}


\begin{table}
\caption{
\label{spinval2}
Comparison of $\tau_{{\rm sp}}$ of the heavy-hole exciton 
due to the deformation potential interaction for 
different configurations of GaAs/Al$_{0.3}$Ga$_{0.7}$As systems at 
$|{\vec{K}}|$ = 1 $\times 10^7$  m$^{-1}$ with $T_{\rm lat} = 4.2$~K,
$T_{\rm ex}=20$~K. 
Other parameters
used are listed in table\ \ref{parameter}.}
{\footnotesize
\begin{tabular}{@{}lllllll}
\br
&Radius&Thickness&Dimensionality
&$\tau_{{\rm sp}}$(ps)&$\tau_{{\rm sp}}$(ps)
\\ 
&R (\AA)& L (\AA)&$\alpha$ &HH$\rightarrow$HH (Scat) &HH$\rightarrow$e+h (Decay)
\\
\mr
quantum wire&50&-&1.31&450&$1 \times 10^3$
\\ 
quantum wire&100&-&1.93&82&150
\\ 
quantum disk &30&50&
0.85&$2.7 \times 10^3$&$1.1 \times 10^4$
\\ 
quantum disk &100&100&1.72&135&223
\\
quantum disk &300&150&2.24&50&83
\\
\br
\end{tabular}}
\end{table}


\section{Appendix}
\label{sec:appen}

The hole wavefunction can be written as a superposition of
four states of the valence band top (labelled by $m = \pm{3\over2}, \pm\half$), 
characterized by the Luttinger parameters $\gamma_1,\gamma_2$ and $\gamma_3$
which appear in the $4\times 4$ Luttinger Hamiltonian  \cite{lutt}:
\beq
\label{Hh}
\hat H_c
= \bordermatrix{& & &  & \cr & H_{{\rm hh}} & c & b & 0 \cr
     & c^* & H_{{\rm lh}} & 0 & -b \cr
     & b^* & 0 & H_{{\rm lh}} & c \cr
     & 0 & -b^* & c^* & H_{{\rm hh}} \cr
     } + V_{{\rm h}}(z_{{\rm h}},L_{{\rm w}}) \; {\mathbb I}_4\, ,
\eeq
where ${\mathbb I}_4$ is the $4\times4$ identity matrix and
where
the heavy-hole and light-hole Hamiltonians 
$H_{{\rm hh}},H_{{\rm lh}}$ and the 
mixing parameters are given by
\begin{eqnarray}
\nonumber
\label{eq:luttH}
H_{{\rm hh}}=-\frac{1}{2m_0} \; \langle p_{z_{{\rm h}}}^2 \rangle 
\left(\gamma_1 - 2 \gamma_2 \right) - \frac{\hbar^2 (k_x^2+k_y^2)}{2m_0}
\left(\gamma_1+\gamma_2\right)
\\
\nonumber
H_{{\rm lh}}= -\frac{1}{2m_0} \; \langle p_{z_{{\rm h}}}^2 \rangle 
\left(\gamma_1+ 2 \gamma_2\right) - \frac{\hbar^2 (k_x^2+k_y^2)}{2m_0}
\left(\gamma_1-\gamma_2\right) 
\\
\nonumber
c(k_x,k_y)=\sqrt{3}\,\gamma_2 \, \frac{\hbar^2}{2 m_0}  (k_x-\rmi k_y)^2
\\
\nonumber
b(k_x,k_y,p_{z_{{\rm h}}})=\sqrt{3}\, \gamma_2 \,
\frac{\hbar}{2m_0}\; \langle p_{z_{{\rm h}}} \rangle 
(k_x-\rmi k_y),
\end{eqnarray}
where $m_0$ is the free-electron mass and 
$V_{{\rm h}}(z_{{\rm h}},L_{{\rm w}})$ is the square well 
potential of thickness $L_{{\rm w}}$ that confines
the holes.
${\bf k_{{\rm h}}}=\left( k_{x_{{\rm h}}},k_{y_{{\rm h}}},k_{z_{{\rm h}}}\right)$
denotes the three-dimensional wavevector of the holes. 
We have also used the axial approximation $\gamma_2 \approx
\gamma_3$ to simplify $c(\vec{k}_{{\rm h}})$ to a single term
and have replaced $k_{z_{{\rm h}}}$ by $-\rmi\partial/\partial z_{{\rm h}}$ 
in the expression for $\langle p_{z_{{\rm h}}}^2 \rangle$.
The material parameters $\gamma_1,\gamma_2,\gamma_3$ are related to the
inverse effective masses of electrons and holes. In the 
$k_{z_{\rm h}}$ direction the  heavy-hole, light-hole masses are given
by, respectively,
\[
m_{\rm hh}={1\over \gamma_1-2\gamma_2},\qquad
m_{\rm lh}={1\over\gamma_1+2\gamma_2}.
\]
The transverse masses of the charge carriers are given by the coefficients
of the $k_x^2+k_y^2$ terms. Consequently the light-hole transverse mass
is larger than the heavy-hole transverse mass.

The hole wavefunctions in a quantum well are obtained by solving:
\beq
\fl
\label{eq:holenergy}
\sum_{\sigma_z^\prime} \; \left [
 \langle \sigma_z | {\hat H_c}| \sigma_z^\prime \rangle 
\, + \, V_{{\rm h}}(z_{{\rm h}},L_{{\rm w}}) \delta_{\sigma_z \, \sigma_z^\prime} \right ]\, 
\xi_{k_{{\rm h}}}^{\sigma_z}(z_{{\rm h}})\; 
 = E_{{\rm h}}(\vec{k}_{{\rm h}}) \, \xi_{k_{{\rm h}}}^{\sigma_z}(z_{{\rm h}})\;
\eeq
where the hole energy $E_{{\rm h}}(\vec{k}_{{\rm h}})$ corresponds to
a heavy-hole or light-hole, depending on the mass
used to model the profile of the 
potential function $V_{{\rm h}}(z_{{\rm h}})$. The summation
is over the four spin projections $\sigma_z = \pm\case32, \pm\case12$ of 
the hole spin, with $\sigma$ aligned along the 
quantization axis.
When $b=c=0$ the Hamiltonian $\hat H_c$ splits into two independent
Hamiltonians, one each for the heavy-hole ($\sigma_z=\pm{3\over2}$)
and light-hole ($\sigma_z=\pm{1\over2}$) cases.


\end{document}